\documentclass[lettersize,journal]{IEEEtran}

\IEEEoverridecommandlockouts

\usepackage{amsmath,amsfonts}

\usepackage{cite}

\usepackage{graphicx}
\usepackage{textcomp}
\usepackage{xcolor}
\def\BibTeX{{\rm B\kern-.05em{\sc i\kern-.025em b}\kern-.08em
    T\kern-.1667em\lower.7ex\hbox{E}\kern-.125emX}}
\usepackage[shortlabels]{enumitem}
\usepackage[normalem]{ulem}

\usepackage[spaces,hyphens]{xurl} 
\usepackage[colorlinks,allcolors=blue]{hyperref}

\usepackage[utf8]{inputenc}
\usepackage{pgfplots}
\pgfplotsset{
compat=1.3,
legend style={font=\footnotesize, fill opacity=0.7,  draw opacity=1, text opacity=1, draw=white!15!black, legend cell align=left, align=left}, 
width=6.5cm, 
every axis/.append style={label style={font=\small},},
yminorticks=false,
xminorticks=false,
title style={font=\small},
tick style={color=black},
tick label style={font=\small},
grid style={line width=.1pt, draw=gray!20},
major grid style={line width=.1pt,draw=gray!20},
}
\DeclareUnicodeCharacter{2212}{−}
\usepgfplotslibrary{groupplots,dateplot}
\usetikzlibrary{patterns,shapes.arrows}
\pgfplotsset{compat=newest}

\usepackage[font=footnotesize]{caption}
\usepackage[font=footnotesize]{subcaption}

\usepackage{array}
\usepackage{textcomp}
\usepackage{stfloats}
\usepackage{url}
\usepackage{verbatim}

\usepackage{tabularx}
\usepackage{booktabs}
\usepackage{colortbl}

\usepackage{algorithmic}
\usepackage[ruled,vlined,linesnumbered]{algorithm2e}
\SetKwComment{Comment}{/* }{ */}

\usepackage{amssymb,amsmath,amsfonts,amsthm,bm,mathtools,cuted,bbold}

\newtheorem{remark}{Remark}

\newtheorem{corollary}{Corollary}

\newtheorem{assumption}{Assumption}

\usepackage[cmintegrals]{newtxmath}

\newcommand{\mc}[1]{\mathcal{#1}}   
\DeclareMathOperator*{\argmax}{arg\,max}    
\DeclareMathOperator*{\argmin}{arg\,min}    
\DeclareMathOperator{\sinc}{sinc}

\newcommand{\DL}{^{\rm \scriptscriptstyle {DL}}} 
\newcommand{\UL}{^{\rm \scriptscriptstyle {UL}}}

\newcommand{\ltwonorm}[1]{\left\lVert#1\right\rVert_2} 

\usepackage[acronym,,nohypertypes={acronym,notation}]{glossaries}
\newacronym{3d}{3D}{three dimensional}

\newacronym{arq}{ARQ}{automatic repeat request}
\newacronym[plural=ACKs]{ack}{ACK}{acknowledgment}
\newacronym{aoa}{AoA}{angle of arrival}
\newacronym{awgn}{AWGN}{additive white Gaussian noise}
\newacronym{aod}{AoD}{angle of departure}

\newacronym{rat}{RAT}{reflected-angular training}
\newacronym[plural=APs, firstplural=access points (APs)]{ap}{AP}{access point}
\newacronym{b5g}{B5G}{Beyond-5G}
\newacronym[plural=BSs, firstplural=base stations (BSs)]{bs}{BS}{base station}

\newacronym{cc}{CC}{control channel}
\newacronym{ce}{CE}{configuration estimation}
\newacronym{chest}{CHEST}{channel estimation}
\newacronym{csi}{CSI}{channel state information}
\newacronym{cdf}{cdf}{cumulative distribution function}
\newacronym{crc}{CRC}{cyclic redundancy check}
\newacronym{crlb}{CRLB}{Cram\'er-Rao lower bound}

\newacronym{dc}{DC}{direct current}
\newacronym{dsp}{DSP}{digital signal processing}
\newacronym{dl}{DL}{downlink}
\newacronym{dlc}{DLC}{data link control}
\newacronym{doa}{DoA}{direction-of-arrival}
\newacronym{emf}{EMF}{electromagnetic field}
\newacronym{em}{EM}{electromagnetic}
\newacronym{fp}{FP}{fractional program}
\newacronym{glrt}{GLRT}{generalized likelihood ratio test}
\newacronym[plural=HRISs, firstplural=Hybrid Reconfigurable Intelligent Surfaces (HRISs)]{hris}{HRIS}{hybrid reconfigurable intelligent surface}
\newacronym[first=i.i.d.]{iid}{i.i.d.}{independent and identically distributed}
\newacronym{ios}{IoS}{Internet-of-Surfaces}
\newacronym{iot}{IoT}{Internet-of-Things}
\newacronym[plural=KPIs, firstplural=key performance indicators (KPIs)]{kpi}{KPI}{key performance indicator}
\newacronym{ls}{LS}{least-squares}
\newacronym{lf}{LF}{low frequency}
\newacronym{los}{LoS}{line-of-sight}

\newacronym{mac}{MAC}{medium access control}
\newacronym{mimo}{MIMO}{multiple-input multiple-output}
\newacronym{mmimo}{M-MIMO}{massive MIMO}
\newacronym{miso}{MISO}{multiple-input single-output}
\newacronym{ml}{ML}{machine learning}
\newacronym{mle}{ML}{maximum-likelihood estimator}
\newacronym{mmse}{MMSE}{minimum mean squared error}
\newacronym{mmtc}{mMTC}{massive machine-type communications}
\newacronym{mrc}{MRC}{maximum-ratio combining}
\newacronym{mse}{MSE}{mean-squared error}
\newacronym{nlos}{NLoS}{non-line-of-sight}

\newacronym{phy}{PHY}{physical}
\newacronym[plural=PDFs]{pdf}{PDF}{probability distribution function}
\newacronym{pla}{PLA}{planar linear array}
\newacronym{pap}{P\&P}{plug-and-play}
\newacronym{ppp}{PPP}{Poisson point process}

\newacronym{ra}{RA}{random access}
\newacronym{rap}{RAP}{random access procedure}
\newacronym[plural=RISs, firstplural=reconfigurable intelligent surfaces (RISs), first=RIS]{ris}{RIS}{reconfigurable intelligent surface}
\newacronym{rf}{RF}{radio frequency}
\newacronym{rmse}{RMSE}{root-mean-square error}
\newacronym{rss}{RSS}{received signal strength}

\newacronym{saloha}{S-ALOHA}{slotted ALOHA}
\newacronym{se}{SE}{squared error}
\newacronym{sdp}{SDP}{semidefinite programming}
\newacronym{sdr}{SDR}{semidefinite relaxation}
\newacronym{sic}{SIC}{successive interference cancellation}
\newacronym{sinr}{SINR}{signal-to-interference-plus-noise ratio}
\newacronym{smse}{SMSE}{sum mean squared error}
\newacronym{sdma}{SDMA}{space-division multiple-access}
\newacronym{snr}{SNR}{signal-to-noise ratio}
\newacronym{soa}{SoA}{state-of-the-art}
\newacronym{sre}{SRE}{smart radio environment}

\newacronym{toa}{ToA}{time-of-arrival}

\newacronym{tdm}{TDM}{time-division multiplexing}
\newacronym{tdma}{TDMA}{time-division multiple access}
\newacronym{tdd}{TDD}{time-division duplex}
\newacronym{tem}{TEM}{transverse electromagnetic mode}

\newacronym[firstplural=users equipment (UEs), plural=UEs]{ue}{UE}{user equipment}
\newacronym{ul}{UL}{uplink}
\newacronym{ula}{ULA}{uniform linear array}
\newacronym{upa}{UPA}{uniform planar array}
\newacronym{uatf}{UatF}{use-and-then-forget}

\newacronym{mr}{MR}{Maximal-ratio}

\definecolor{gold}{rgb}{0.85,.66,0}
\definecolor{amaranth}{rgb}{0.9, 0.17, 0.31}

\begin{document}

\bstctlcite{IEEEexample:BSTcontrol} 

\title{
   Random Access Protocol with Channel Oracle Enabled by a Reconfigurable Intelligent Surface
}

\author{
    Victor~Croisfelt,~\IEEEmembership{Graduate Student Member,~IEEE,}
    Fabio~Saggese,~\IEEEmembership{Member,~IEEE,}
    Israel~Leyva-Mayorga,~\IEEEmembership{Member,~IEEE,}
    Rados\l{}aw~Kotaba, 
    Gabriele~Gradoni,~\IEEEmembership{{Member,~IEEE,}}
    and Petar~Popovski,~\IEEEmembership{Fellow,~IEEE}\\
    \vspace{5mm}
    {\small
    \textbf{NOTE:} This work has been submitted to the IEEE TWC for possible publication.\\
    Copyright may be transferred without notice, after which this version may no longer be accessible.
    }
    \thanks{
        V. Croisfelt, F. Saggese, I. Leyva-Mayorga, R. Kotaba, and P. Popovski are with the Connectivity Section of the Department of Electronic Systems, Aalborg University, Aalborg, Denmark (e-mail: \{vcr,fasa,ilm,rak,petarp\}@es.aau.dk). G. Gradoni is with the Department of Electrical and Electronics Engineering, University of Nottingham, Nottingham, United Kingdom (e-mail: Gabriele.Gradoni@nottingham.ac.uk).
    }
}

\maketitle

\begin{abstract}
    The widespread adoption of Reconfigurable Intelligent Surfaces (RISs) in future practical wireless systems is critically dependent on the integration of the RIS into higher-layer protocols beyond the physical (PHY) one, an issue that has received minimal attention in the research literature. In light of this, we consider a classical random access (RA) problem, where uncoordinated users' equipment (UEs) transmit sporadically to an access point (AP). Differently from previous works, we ponder how a RIS can be integrated into the design of new medium access control (MAC) layer protocols to solve such a problem. We consider that the AP is able to control a RIS to change how its reflective elements are configured, namely, the RIS configurations. Thus, the RIS can be opportunistically controlled to favor the transmission of some of the UEs without the need to explicitly perform channel estimation (CHEST). We embrace this observation and propose a RIS-assisted RA protocol comprised of two modules: Channel Oracle and Access. During channel oracle, the UEs learn how the RIS configurations affect their channel conditions. During the access, the UEs tailor their access policies using the channel oracle knowledge. Our proposed RIS-assisted protocol is able to increase the expected throughput by approximately 60\% in comparison to the slotted ALOHA (S-ALOHA) protocol.
\end{abstract}


\IEEEpeerreviewmaketitle

\section{Introduction}
\IEEEPARstart{R}{econfigurable} Intelligent Surfaces (RISs) are low-power wireless system elements that can shape the radio waves to enable smart radio environments~\cite{Huang2019,strinati2021rise6g}. The majority of the literature on \gls{ris}-assisted communication systems has predominantly focused on the \gls{phy} layer aspects, including the modeling of the related electromagnetic phenomena~\cite{Pizzo2022,Pizzo2022a}. Many papers have shown potential benefits in terms of spectral and energy efficiencies of \gls{ris}-assisted systems~\cite{bjornson2021signalprocessing}, which are based on the development of optimization methods for the \gls{ris} alone or jointly with the \gls{ap} precoding~\cite{Ross2021, jamali2022low, Mursia2020}. Because of this, several works have focused on the design and evaluation of \gls{chest} procedures in the presence of \gls{ris}~\cite{Wei2021ce}. A particular effort has been made to reduce the overhead of \gls{chest}, which can be very high due to the high number of elements that compose the \gls{ris}~\cite{yuan2022tensor, yuan2021frequency, wang2020compressed}. However, in general, less attention has been paid to integrating the \gls{ris} into protocols at higher layers to leverage the \gls{phy} benefits provided by the \gls{ris}. 

An interesting first problem for the integration of the \gls{ris} into higher layers is to consider the \gls{ra} problem or the ``free-for-all'' multiaccess communication~\cite{Bertsekas1996,Popovski2020}, where individual resource allocation and coordination among \glspl{ue} are not feasible. The interesting questions here are $i$) How to coordinate transmissions to avoid collisions so that exactly one \gls{ue} is transmitting for a given time period? and $ii$) When and how to retransmit packets when collisions occur? Another complicating factor is that the \gls{ap} just know the area where the \glspl{ue} are placed. Conventionally, these issues are addressed at the \gls{mac} layer, which implements protocols to allocate the multiaccess medium among \glspl{ue}~\cite{Bertsekas1996,Popovski2020}. In particular, we refer to an \gls{ra} protocol as a distributed algorithm whose common objective is to solve the aforementioned questions. 

Legacy \gls{mac} protocols are often designed by taking only \glspl{ap} and \glspl{ue} as communication nodes, such as in \gls{saloha}~\cite{Bertsekas1996,Popovski2020}. However, by introducing the \gls{ris} into the environment, the shared channel is definitely affected. To see this, consider a typical wireless multiaccess channel, where the received signal at the \gls{ap} is the sum of attenuated transmitted signals from a set of \glspl{ue}, with the signals being corrupted by distortion, delay, and noise~\cite{Bertsekas1996}. In a \gls{ris}-assisted wireless multiaccess channel, the signals of some of these \glspl{ue} can be \textit{intentionally} favored over others due to the \gls{ris}' reflecting capabilities without the need to explicitly perform \gls{chest}. In particular, we are interested in the case that an \gls{ap} controls a \gls{ris} by being able to change how the reflective elements that compose the RIS are configured, namely, the \gls{ris} configurations. Each of the \gls{ris} configurations directs an incoming wave in a specific direction, called the reflected angular direction or reflection angle. In this paper, we propose a \gls{ris}-assisted \gls{ra} protocol that distills the above insight by separating the \gls{ue} transmissions in time based on their geographical location and the design of \gls{ris} configurations, allowing the \glspl{ue} to opportunistically determine the adequate time to transmit by running an algorithm locally.

\subsection{Related Works}
For \gls{ris}-assisted wireless systems, there is a gap in the literature regarding the \gls{ra} and related problems. The authors of~\cite{Mursia2020,Cao2022} present designs for \gls{mac} protocols that integrate \glspl{ris} for multi-user communications. These works address the multiaccess problem based on the ``perfectly scheduled" approach~\cite{Bertsekas1996}, where the \glspl{ap} already know the \glspl{ue} because they have been scheduled somehow. This is conceptually different from the \gls{ra} problem being addressed here, where the latter is more challenging since few things are known among communication nodes. For example, the \gls{ra} problem precedes the \gls{chest}, since scheduling had not been realized yet. In this regard, a closely related work is~\cite{Shao2021bayesiantensor}, where the authors consider the activity detection problem for unsourced \gls{ra} by using the \gls{ris} to improve the channel quality and control channel sparsity. However, this work does not clarify how to adequately integrate the \gls{ris} into the protocol design and uses the \gls{phy} controlling capabilities of \gls{ris} only as an artifact to improve channel sparsity under very specific operating conditions, such as the use of millimeter waves. In addition, the authors in~\cite{Laue2022detection} propose an activity detection algorithm that optimizes the \gls{ris} configuration to obtain optimal detection probability. However, the procedure relies on a partial \gls{chest}. In~\cite{Kherani2022randomaccess}, the authors analyze the performance of a \gls{ris}-assisted \gls{ra} using \gls{sic} for uncoordinated transmission attempts from two transmitters demonstrating that the \gls{ris} can help achieve better performance due to increased \gls{snr}. Nevertheless, the work also lacks the systematical aspects behind a protocol. To address these problems, in~\cite{Croisfelt2022}, we have proposed a proof-of-concept of the \gls{ris}-assisted \gls{ra} protocol that is going to be shown here. In there, we have shown substantial gains in integrating the \gls{ris} into the \gls{ra} protocol design. Still, we have omitted several engineering details of the protocol, important for the overall system design.

\subsection{Contributions}
Our proposed \gls{ris}-assisted \gls{ra} protocol is based on the observation that the \gls{ap} can control the \gls{ris} to intentionally favor the transmission of some of the \glspl{ue} without the need of \gls{chest}. Naturally, one might think that because of this the \gls{ris} might act as a mediator or coordinator in a completely uncoordinated environment bringing some light into the darkness. However, for this coordination to be possible, we had the idea that \glspl{ue} need to be aware of when they are being favored. That is, they need to know how their channels change w.r.t. the reflected angular space spanned by the \gls{ris} configurations. Thus, our protocol comprises two modules: A. Channel Oracle and B. Access. During the channel oracle, each \gls{ue} learns when and how its channel is favored by the \gls{ris} considering its current positions. During access, each \gls{ue} exploits the channel oracle knowledge to come up with a tailored access policy that can be designed to define when to retransmit packets and to avoid collisions. Our proposed \gls{ris}-assisted \gls{ra} protocol has several compelling advantages: it improves overall \gls{mac} performance in comparison with legacy protocols, it encourages the installation of \glspl{ris} instead of new \glspl{ap}, which could end up to be cheaper and more energy efficient, and it provides the \gls{ap} with relevant information to conduct other operations that start after the \gls{ra}. For example, it can alleviate the computational complexity of \gls{chest}, since the access policies, if properly designed, tend to reveal information about where \glspl{ue} are located and their channel conditions.

The remainder of the paper is organized as follows. In Section~\ref{sec:system-model}, we introduce our system model by setting up all that is need for the presentation of the protocol. In Section~\ref{sec:protocol} the protocol is presented more broadly. We give particular details on how to design the Channel Oracle and Access modules in Sections \ref{sec:channel-oracle} and \ref{sec:access}, respectively. In Section \ref{sec:practical}, we discuss several practical details of the protocol and how they can be extended to some other system models and set of assumptions. Finally, we numerically evaluate our protocol in Section~\ref{sec:results}, whereas Section~\ref{sec:conclusions} draws our main conclusion.

\noindent \textbf{Notations.} The set of positive integers, positive real, real, and complex numbers are denoted by $\mathbb{Z}_{+}$, $\mathbb{R}_{+}$, $\mathbb{R}$, and $\mathbb{C}$, respectively. Integer sets are denoted by calligraphic letters $\mc{A}=\{0,1,\dots,A-1\}$ with cardinality $|\mc{A}|=A$. The circularly-symmetric complex Gaussian distribution is $\mc{N}_{\mathbb{C}}(\mu,\sigma^2)$ w/ mean $\mu$ and variance $\sigma^2$. Lower and upper case boldface letters denote column vectors $\mathbf{x}$ and matrices $\mathbf{A}$, respectively. The identity matrix of size $N$ is $\mathbf{I}_N$ and $\mathbf{0}$ is a vector of zeros of arbitrary size. Euclidean norm is $\lVert\mathbf{x}\rVert_2$. Superscript $(\cdot)^*$ denotes complex conjugate. The $\arg\max(\cdot)$ function returns the index of the maximum element of a vector, while $\mathrm{med}\{\cdot\}$ and $\mathrm{max}\{\cdot\}$ denote the median and the maximum operator over a set, respectively.

\section{System Model}\label{sec:system-model}
\begin{figure}[ht]
    \centering
    \includegraphics[trim={0.8cm 3.4cm 0.0cm 0cm}, clip]{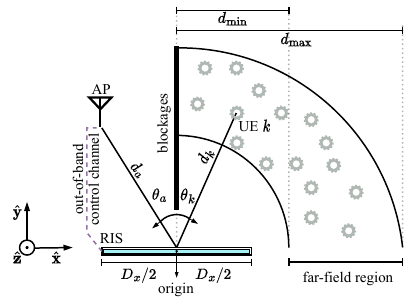}
    \caption{Two-dimensional view of the considered system setup assisted by a \gls{ris} of negligible thickness.}
    \label{fig:system-setup}
\end{figure}

Consider the wireless local area network depicted in Fig.~\ref{fig:system-setup}, which is comprised of one single-antenna \gls{ap}, one \gls{ris}, and multiple $K\in\mathbb{Z}_+$ single-antenna \glspl{ue}, which are indexed by the set $\mathcal{K}$. We assume that the \gls{ap} does not have any prior knowledge of the \glspl{ue}. This scenario may correspond to an industrial installation, where many sensors and actuators with wireless communication capabilities communicate with an \gls{ap} of a network infrastructure. For example, the \gls{ap} can be deployed outside, while the \glspl{ue} are located within an industrial shed. However, due to possible blockages, such as machines and walls, the \gls{los} paths among the \gls{ap} and the \glspl{ue} are blocked. In this context, the network operator may decide to install a \gls{ris} to improve the quality of the communication among the \gls{ap} and the \glspl{ue}, instead of installing a new \gls{ap} within the shed. Note that it is reasonable to assume only one \gls{ris} since we are dealing with a very controlled application scenario. However, multiple \glspl{ris} can be considered, where each \gls{ris} helps the communication of different sets of geographically isolated \glspl{ue}; \emph{e.g.}, located in several sheds.\footnote{Multiple \glspl{ris} can also be used in a cooperative way to serve a single set of \glspl{ue}. A discussion on how to extend the proposed framework to multiple \glspl{ris} is given in Sect.~\ref{sec:practical}.} Moreover, we consider that the \gls{ap} controls the operation of the \gls{ris} through a dedicated out-of-band \gls{cc}, \emph{i.e.}, controlling the \gls{ris} does not interfere with the wireless signals exchanged among the \gls{ap} and \glspl{ue}~\cite{bjornson2021signalprocessing}. For convenience, we also assume that the \gls{cc} is error-free with the command messages sent by the \gls{ap} being perfectly interpreted by the \gls{ris} controller (RIS-C) at the \gls{ris}' side.

To keep the complexity of the analysis at a minimum while showing the working principles of the proposed \gls{ris}-assisted \gls{ra} protocol, we will analyze the performance of the system under the following assumptions.

\begin{assumption}[Ideal \gls{ris}] \label{assu:ideal}
    The considered \gls{ris} is assumed to have an ideal hardware able to induce a stable phase shift on the incident wave without affecting its amplitude, and with negligible mutual coupling among reflecting elements.
\end{assumption}

\begin{assumption}[\glspl{ue} Positioning] \label{assu:position}
    The \gls{ap} and the \glspl{ue} are located on the $x$-$y$ plane having $z$-coordinate equal to 0, \emph{i.e.}, they lay on the same plane of the center of the \gls{ris}.
\end{assumption}

In a real implementation, the \gls{ris}' reflecting elements, or simply elements, generate an attenuation loss which is a function of the selected phase shift~\cite{Cai2020practical, Li2021practical}. Moreover, each element can only induce a phase shift whose value comes from a finite set~\cite{Ross2021} due to hardware constraints. Finally, inter-element coupling effects might exist~\cite{Qian2021coupling}, while usually assumed negligible~\cite{bjornson2021signalprocessing, tang2020wireless}. Thus, the results provided in this paper can be seen as an upper bound in terms of performance since attenuation, finite precision, and mutual coupling would result in performance losses. On the other hand, constraining the position of the \glspl{ue} on the same plane allows us to present the protocol structure analyzing the problem as a function of the incident angle only, \emph{i.e.}, in a single angular dimension. Thus, Assumptions~\ref{assu:ideal} and~\ref{assu:position} yield a simple analytical model that can be exploited to clearly present the main, new ideas behind the proposed \gls{ris}-assisted \gls{ra} protocol. We discuss possible directions on how to extend the protocol to other scenarios with a different set of assumptions in Sect.~\ref{sec:practical}.

\subsection{Geometry}\label{sec:geometry}
As illustrated in Fig.~\ref{fig:system-setup}, the \gls{ris} is positioned in such a way that it can reflect signals transmitted by the \gls{ap} into the area where the \glspl{ue} are located and vice versa.\footnote{The scenario can be straightforwardly extended to different geometries, as long as the direct path between \gls{ap} and \gls{ue} is blocked.} The center of the \gls{ris} is the origin of our coordinate system. The \gls{ris} is formed by $M_x\in\mathbb{Z}_{+}$ and $M_z\in\mathbb{Z}_{+}$ elements arranged in a planar array over the $x$- and $z$-dimensions, with index sets $\mc{M}_x$ and $\mc{M}_z$, respectively, and totaling $M=M_xM_z$ elements. Each wavelength-scale element is realized as a metalized layer on a grounded substrate and has an area of $d_x d_z$ with $d_x,d_z\in\mathbb{R}_+$ and $d_x,d_z\leq\lambda$, where $\lambda$ is the wavelength of the carrier signal. The dimensions of the \gls{ris} are $D_x=M_xd_x$ and $D_z=M_zd_z$. Following Assumption~\ref{assu:position}, we denote as $\theta_{a}\in[0,{\pi}/{2}]$ the angle between the line normal to the origin and the \gls{ap}, and as $\theta_{k}\in[0,{\pi}/{2}]$ the equivalent angle w.r.t. the $k$-th \gls{ue}, $\forall k\in\mathcal{K}$. The corresponding distances are denoted as $d_{a}$ and $d_k\in[d_{\min},d_{\max}]$, where $d_{\min}$ and $d_{\max}$ denote the minimum and maximum distances, respectively. The minimum distance is set as $d_{\min}=\frac{2}{\lambda}\max(D_x^2,D_z^2)$ so that we can analyze the electromagnetic signals reflected by the \gls{ris} in the \emph{far-field regime}~\cite{Balanis2012eletro}, assuring the assumption of plane-wave propagation. The maximum distance {$d_{\max}$} stipulates the lower boundaries of the \gls{snr} range, \emph{i.e.}, it is set as the maximum distance the \gls{ris} is able to provide a sufficient good \gls{snr} at the \glspl{ue}.

\subsection{RIS Configurations}\label{sec:system-model:phase-shift-configurations}
According to Assumption~\ref{assu:ideal}, we let $\phi_{m,m'}\in[0,2\pi],\,\forall m\in\mc{M}_x,\ \forall m'\in\mc{M}_z$ denote the phase shift impressed by the $(m,m')$-th element. Without loss of generality, we assume a transverse electromagnetic mode propagation~\cite{Balanis2012eletro}, where the electromagnetic waves propagate within the plane perpendicular to the $z$-axis~\cite{Ross2021}. Consequently, and in agreement with Assumption~\ref{assu:position}, the \gls{ris}' elements with the same index over the $z$-dimension impress the same phase shift, to maximize the energy on the $x-y$ plane. Hence, for notation convenience, the dependency with the $z$-dimension can be dropped as $\phi_{m,m'}=\phi_{m,m''}=\phi_{m},\,\forall m',m''\in\mc{M}_z$. We then refer to a vector of phase shifts $\boldsymbol{\phi}\in[0,2\pi]^{M_x M_z}$ as a \emph{\gls{ris} configuration} denoting a particular way in which the reflective elements of the \gls{ris} are configured to scatter the incoming wave. In our context, we consider that the \gls{ris} has a finite number $N\in\mathbb{Z}_+$ of predefined configurations collected in the so-called \emph{\gls{ris} configuration codebook} $\Phi=\{\boldsymbol{\phi}[n]\}_{n\in\mc{N}}$, where $\mathcal{N}$ is the index set and $[n]$ is used for indexing as a notation convenience based on classical signal processing literature~\cite{Proakis2006}. \emph{We note that the \gls{ris} configuration codebook $\Phi$ can be comprised of many subsets $\Phi_i$, where each configuration codebook can be designed for a particular task}. During the network setup, {the configuration codebook $\Phi$ is loaded into the \gls{ap} and \gls{ris} hardware. {During the network operations}, the \gls{ap} controls the behavior of the \gls{ris} by referring to configurations of $\Phi$. 

\noindent\textbf{From Reflection Angles to Configurations.} Each configuration can be related to the angle of the resulting reflected wave, denoted as $\theta\DL_r, \theta\UL_r \in[0,{\pi}/{2}]$, for \gls{dl} and \gls{ul} directions, respectively. Mathematically, let $h:\Theta\mapsto\Phi$ be the bijective mapping among reflection angles and configurations. We now analyze how to design a configuration codebook having in view the reflection angles and the facts that the \gls{ap} controls the \gls{ris} and does not have prior knowledge of the \glspl{ue}' positions when wishing to enable \gls{ra} functionalities. In the \gls{dl}, suppose that the \gls{ap} sends a signal toward the \gls{ris}, whose incoming direction is $\theta_a$. The \gls{ap} can control the \gls{ris} to reflect the incoming wave toward a desired \gls{dl} angular direction $\theta\DL_r$. In our context, the \gls{ap} wishes to \emph{sense} a particular angular direction looking for \glspl{ue}, that is, it would like to choose a \gls{dl} reflection direction $\theta\DL_r$ that matches with a $\theta_k,\,k\in\mathcal{K}$. In other words, the reflection angle $\theta\DL_r$ is chosen blindly by the \gls{ap} with the intention of matching with the angular position of a \gls{ue} since the \gls{ap} does not know the \glspl{ue}' angular positions $\theta_k$. Now, in the \gls{ul}, suppose that a single or multiple \glspl{ue} transmit, meaning that the incoming signal(s) from the \gls{ris}'s standpoint comes from random directions $\theta_k$. Since the \gls{ap} controls the \gls{ris}, the best the \gls{ap} can guess is that the incoming waves are coming from the reflection directions $\theta\DL_r$'s used during the \gls{dl} instead of $\theta_k$'s. This consequently interlaces both the \gls{dl} and \gls{ul} phases. The \gls{ap} can then control the \gls{ris} to reflect the incoming waves toward a desired \gls{ul} angular direction $\theta\UL_r$. Naturally, the \gls{ap} controls the \gls{ris} to reflect the scattered incoming waves toward itself, \emph{i.e.}, $\theta\UL_r=\theta_a$. Based on this, the configuration design from reflection angles to configurations can be written as~\cite{Balanis2012antenna}:\footnote{This design can also be motivated by the Generalized Snell's Law~\cite{Ozdogan2020}.}
\begin{align}
    \phi\DL_m &= \omega \, d_x \, (m+1) (\sin\theta_a - \sin\theta\DL_r), \text{ and} \nonumber \\
    \phi\UL_m &= \omega \, d_x \, (m+1) (\sin\theta\DL_r - \sin\theta_a),
    \label{eq:system-model:optimal-config}  
\end{align} 
 where $\phi\DL_m$ and $\phi\UL_m$ are the phase shifts impressed at the $m$-th element under \gls{dl} and \gls{ul} transmissions, respectively, $\forall m\in\mc{M}_x$. Given that we have $N$ configurations, we can denote the \gls{dl} and \gls{ul} the bijective mappings as $\boldsymbol{\phi}\DL[n]=h\DL(\theta\DL_r[n])$ and $\boldsymbol{\phi}\UL[n]=h\DL(\theta\UL_r[n])$, respectively. Note that from the general case above $\phi\DL_m=-\phi\UL_m$ since $\theta\UL_r=\theta_a$, meaning that the corresponding mappings are related as $h\DL(\theta\DL_r)=(h\UL(\theta\UL_r))^{*}$. Thus, since the \gls{dl} configuration is just the complex conjugate of the \gls{ul} one due to channel reciprocity, the \gls{ap} and \gls{ris} can just agree on a "unidirectional" configuration codebook $\Phi$ and the \gls{ap} signalizes either if the configuration should be complexly conjugated, indicating \gls{ul}, or not for \gls{dl}.

\subsection{RIS-Assisted Slotted Multiaccess}\label{sec:system-model:mac:frame}
Consider a standard \emph{frame} structure of \gls{dlc}, where each frame represents the data stream that a single \gls{ue} wants to transmit toward the \gls{ap}.
A frame has fixed duration $T_{\rm F}\in\mathbb{R}_+$, comprises a \gls{crc} for error detection, while stop-and-wait \gls{arq} is used for error correction. For \gls{phy} transmission, this frame is sliced into transmission packets. Assume a slotted system for transmission over the \gls{phy} layer, where all transmitted packets have the same duration $T_{\rm s}\in\mathbb{R}_+$ and each packet requires a one-time unit or \emph{slot} for transmission. Thus, the overall frame duration is $T_{\rm F}=N_{\rm F}T_{s}$, where $N_{\rm F}\in\mathbb{Z}_+$ is the number of slots in a frame. Moreover, we consider that each slot is further sliced into $L\in\mathbb{Z}_{+}$ samples or symbols with fixed symbol duration $T_{\rm symb}\in\mathbb{R}_{+}$; hence, $T_{s}=LT_{\rm symb}$. Conventionally, the multiaccess problem culminates in coordinating the use of the slotted \gls{phy} channel so to avoid collisions among \glspl{ue} on a slot basis. However, now we have a \emph{controllable channel} due to the \gls{ris}, leading to new approaches to attain some coordination among \glspl{ue}. We are interested in studying these new possibilities and for this, we assume the following.

\begin{assumption}[Slotted Configuration Change]
    The \gls{ap} can control the \gls{ris} to change its configurations on a slot basis, where a single configuration can be changed per slot. Therefore, at the beginning of each slot, the \gls{ap} sends a configuration change command to the \gls{ris}. After the \gls{ris} receives this command, it needs a certain time to physically load the new configuration. We let $T_{\rm sw}\in\mathbb{R}_{+}$ be the switching time, \emph{i.e.}, the time the \gls{ris} requires to switch to a new configuration. The \gls{ap} remains silent during $T_{\rm sw}$ in the \gls{dl} and ignores any signal received during the switching time in the \gls{ul}. Note that $T_{\rm sw}$ can also incorporate the time spent on sending command messages over the \gls{cc}.
    \label{assu:slotted-config}
\end{assumption}

\subsection{Physical Channel Model} \label{sec:system-model:channel-model}
Assume dominant \gls{los} paths for \gls{ap} to \gls{ris} and \gls{ris} to a \gls{ue} $k$. The \gls{dl} channel coefficient $\zeta_{k}\DL(\boldsymbol{\phi}\DL)\in\mathbb{C}$ is:
\begin{align}
    \zeta_{k}\DL(\boldsymbol{\phi}\DL)&=\sqrt{\beta_{k}\DL}e^{j \omega \psi_{k}}\mathrm{A}_{k}(\boldsymbol{\phi}\DL) \text{ with} \label{eq:system-model:dl-channel-gain}\\
    \beta_{k}\DL&=\dfrac{G_{a} G_{k}}{(4\pi)^2} \left(\dfrac{d_x d_z}{d_{a} d_{k}}\right)^2\cos^2\theta_{a},
\end{align}
where $(\boldsymbol{\phi}\DL)$ denotes the dependency on the phase shifts, $\beta_{k}\DL\in\mathbb{R}_{+}$ is the \gls{dl} pathloss with $G_a$ and $G_k$ being the antenna gain of the \gls{ap} and of the \gls{ue}, respectively. The propagation phase shift $\psi_{k}\in[0,2\pi]$ and the array factor arising from the discretization of the \gls{ris} into a finite number of elements are: 
\begin{align} 
    \psi_{k} = - \left( d_a + d_k - (\sin\theta_a - \sin\theta_k) \frac{M_x+1}{2} d_x \right) \text{ and}\\
    \mathrm{A}_{k}(\boldsymbol{\phi}\DL) = M_z \sum_{m\in\mc{M}_x} e^{ j (\omega d_x (m+1) (\sin\theta_k - \sin\theta_a) + \phi\DL_m)},
    \label{eq:system-model:array-factor}
\end{align} 
where $\phi\DL_m\in[0,2\pi]$ is the $m$-th element of $\boldsymbol{\phi}\DL$. Similarly, the \gls{ul} channel coefficient $\zeta_{k}\UL(\boldsymbol{\phi}\UL)\in\mathbb{C}$ and the corresponding pathloss $\beta_{k}\UL\in\mathbb{R}_{+}$ are  
\begin{align}
    \zeta_{k}\UL(\boldsymbol{\phi}\UL) &=
    {\sqrt{\beta_{k}\UL}}
    e^{- j \omega \psi_{k}} \mathrm{A}_{k}(\boldsymbol{\phi}\UL) \text{, and}\label{eq:system-model:ul-channel-gain}\\
    \beta_{k}\UL &= \dfrac{G_{a} G_{k}}{(4\pi)^2} \left(\dfrac{d_x d_z}{d_{a} d_{k}}\right)^2\cos^2\theta_{k}.  
\end{align}
The interested reader can check more details about the derivation of the above model in \cite{Croisfelt2022}. It is worth pointing out that the antenna array model adopted here can be extended to include the near-field of the RIS through recently proposed plane wave expansion methods~\cite{Pizzo2022a}. Note also that, by substituting~\eqref{eq:system-model:optimal-config} into~\eqref{eq:system-model:array-factor}, the array factor becomes:
\begin{equation} 
    \mathrm{A}_{k}(\boldsymbol{\phi}\DL) \equiv \mathrm{A}_{k}(\theta\DL_r) = M_z \sum_{m\in\mc{M}_x} e^{j \omega d_x (m+1)  (\sin\theta_k - \sin\theta\DL_r) }.
    \label{eq:system-model:phase-shifted-array-factor}
\end{equation}
Observe that $\mathrm{A}_{k}(\boldsymbol{\phi}\UL)=\mathrm{A}^{*}_{k}(\boldsymbol{\phi}\DL)$. As a result, the channel coefficients are also a function of the reflection angles, that is, it is equivalent to say that $\zeta\DL_{k}(\boldsymbol{\phi}\DL)\equiv\zeta\DL_{k}(\theta\DL_r)$. Due to the static or low mobility of the main scenario of interest and controllable scattering effects, the channel coefficients are assumed constant over slots.

\section{Proposed RIS-Assisted Random Access Protocol}\label{sec:protocol}
In this section, we present the proposed \gls{ris}-assisted \gls{ra} protocol illustrated in Fig.~\ref{fig:protocol}. The structure of the protocol is composed of two independent modules: $A$. Channel Oracle and $B$. Access. Inspired by the carrier sensing approach \cite{Bertsekas1996,Popovski2020}, the basic idea of our protocol is that each \gls{ue} can learn a model on how its channel coefficient varies over the reflected angular space spanned by the channel control offered by the \gls{ris}. The role of the channel oracle module is to specify how this learning task occurs in a distributed manner since the control of the \gls{ris} is owned by the \gls{ap} and not by the \glspl{ue}. By using the output of the channel oracle, the access module then specifies how the \glspl{ue} attempt to transmit their packets to the \gls{ap} over the multiaccess channel when considering that the \gls{ap} is unaware of any prior information regarding the \glspl{ue} and again owns the control of the \gls{ris}. We note that the access module depends on the output of the channel oracle module, otherwise, the \glspl{ue} would not benefit from the environment control brought by the \gls{ris}. In this case, the access module could be simply replaced by legacy protocols, such as \gls{saloha}~\cite{Bertsekas1996,Popovski2020}. Moreover, one should keep in mind that the access module is realized much more often than the channel oracle one in practice. More practical details of the protocol can be seen in Section \ref{sec:practical}. Below we give an overview of the protocol modules and introduce performance metrics.

\begin{figure}[t]
    \centering
    \includegraphics[trim={0 2mm 0 0}, clip, width=1\columnwidth]{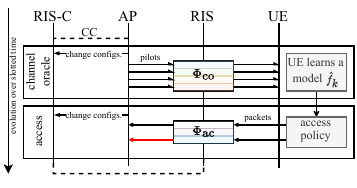}
    \caption{Time diagram of the proposed \gls{ris}-assisted \gls{ra} protocol. The RIS-C denotes the RIS controller, which is connected to the \gls{ap} via the \gls{cc}. Operations occur sequentially in time following the top-down direction. The protocol is comprised of two modules: channel oracle and access. During the channel oracle, the \gls{ap} sends pilots, while the \gls{ris} sweeps through a set of configurations $\Phi_{\rm co}$; this enables the \gls{ue} to learn a model $\hat{\zeta}\DL_k=\hat{f}_k(\theta\DL_r)$. During access, the \gls{ue} exploits $\hat{f}_k$ to design access policies so that they can choose when to send packets, while the \gls{ris} sweeps again through another set of configurations $\Phi_{\rm ac}$. The \textcolor{red}{red arrow} illustrates a \textcolor{red}{collision} since hypothetically another \gls{ue} decided to send a packet during the same access configuration.}
    \label{fig:protocol}
\end{figure}

\subsection{Channel Oracle}\label{sec:protocol:channel-oracle}
Mathematically, the channel oracle at the \gls{ue}'s side consists of each \gls{ue} $k$ learning a model $\hat{f}_k:[0,\pi/2]\mapsto\mathbb{C}$ of the channel such that $\hat{\zeta}\DL_k=\hat{f}_k(\theta\DL_r)$, where the input is a \gls{dl} reflection angle $\theta\DL_r$ and the output is the \gls{dl} estimated channel coefficient $\hat{\zeta}\DL_k$, for $k\in\mathcal{K}$. In order for the \glspl{ue} to be able to learn such a model, they need to obtain some input data by sampling the reflected angular space. This sampling could be done by changing the \gls{ris} configurations over slots. However, the control of the \gls{ris} is exercised by the \gls{ap} and not by the \glspl{ue}, thus characterizing a distributed learning problem. Hence, since the \gls{ap} is in charge of the sampling process, the \gls{ap} needs to meticulously design the minimum set of configurations so that the \glspl{ue} are able to properly obtain the models $\hat{f}_k$ up to a considerably low error bound. Where minimum here comes from the natural desire of reducing any overhead. In order to fulfill this objective, the basic idea is that the \gls{ap} broadcasts pilot signals, while it controls the \gls{ris} to change its configurations. In this way, the \glspl{ue} can obtain input data to learn the model. Note that there is a clear order in this part of the protocol, where the \gls{ap} performs its actions first followed by the \glspl{ue}. Moreover, observe that the channel oracle is performed simultaneously by all the $K$ \glspl{ue} present in the environment with each of them learning its own model.

We now introduce the basic notation related to the channel oracle module. The distributed sampling approach is coordinated by the \gls{ap} and sensed by the \glspl{ue}. The sampling points are specified by the \emph{channel oracle codebook} $\Phi_{\rm co}\equiv\Theta_{\rm co}$, enumerated by $\mc{N}_{\rm co}$ and $|\mc{N}_{\rm co}|={N}_{\rm co}$ being the number of \emph{channel oracle configurations} (samples). We assume that each configuration in $\Phi_{\rm co}$ is loaded for one slot into the \gls{ris}; consequently, the channel oracle takes $T_{\rm co}=N_{\rm co} (L_{\rm co} T_{\rm symb} + T_{\rm sw})$ seconds with $L_{\rm co}$ being the pilot sequence length in a so-called channel oracle slot, which can be adjusted to combat noise. After sampling the reflected angular space over the \gls{dl} direction, each \gls{ue} has collected the pairs $(\hat{\zeta}\DL_k[n],\theta\DL_r[n])_{n\in\mathcal{N}_{\rm co}}$ employed to learn $\hat{f}_k$. Fig.~\ref{fig:system-model:channel-oracle} illustrates the sampling of the \gls{dl} channel gain of two \glspl{ue} located at different positions. The output of the channel oracle module at each device can be obtained by interpolating the sampled points, where the knowledge of the channel oracle codebook is shared with the \glspl{ue} during the network setup.

\begin{figure}[ht]
    \centering
    \vspace{-4mm}
    \input{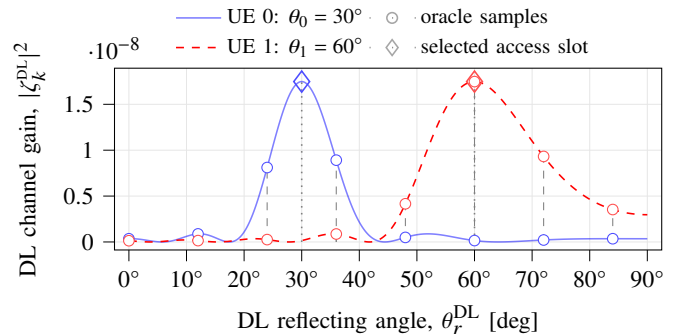}
    \vspace{-6mm}
    \caption{Illustration on how the channel oracle and access modules interact for two \glspl{ue} located in different angular directions. During the channel oracle module, the \gls{ris} sweeps through the channel oracle codebook $\Phi_{\rm co}$ sampling uniformly the channel in the reflected angular space, with $N_{\rm co}=8$ configurations ($\circ$ oracle samples). Each \gls{ue} tries to learn a function $\hat{f}_k$ that predicts the channel gain given the sampled points as information (\textcolor{blue}{blue} and \textcolor{red}{red} solid lines). During the access module, the \glspl{ue} select the most suited access slots from $\Phi_{\rm ac}$ based on $\hat{f}_k$ ($\lozenge$ selected access slot); naturally, the \glspl{ue} wish to choose access slots that are in proximity of the peak of $\hat{f}_k$ to increase the \gls{snr} and the chance of successful transmission. Here, $d_a=5$ m, $\theta_a=45^\circ$ and $d_k=5$ m, $k=\{0,1\}$.}
    \label{fig:system-model:channel-oracle}
\end{figure}

\subsection{Access}\label{sec:protocol:access}
The access module defines the behavior of the \gls{ap} and \glspl{ue} when the latter attempt transmission. We assume that the \gls{ap} establishes that the \glspl{ue} can try to transmit data during an \emph{access period} comprised of $N_{\rm ac}\in\mathbb{Z}_+$ slots enumerated by $\mathcal{N}_{\rm ac}$ and with $N_{\rm ac}\geq N_{\rm F}$. Before the start of this period, consider that a number of $K_a\in\mathbb{Z}_{+}$ (active) \glspl{ue} unpredictably wake up to transmit one data packet fitted in the \gls{dlc} frame structure, with $K_a\ll K$ being unknown to the \gls{ap} and with index set $\mc{K}_a\subset\mathcal{K}$. Eventually, we assume that \glspl{ue} waking up within the access period will wait for the next one to start their transmission. The start of the access period is signaled by an \gls{ap} beacon, and perfect synchronization is assumed. Given that, a \emph{collision} occurs when two or more active \glspl{ue} try to send a packet in a given slot. We further assume immediate slot feedback at the end of each slot, meaning that the \gls{ap} can let the \glspl{ue} know if the packet transmission was either successful, absent or if a collision occurred.\footnote{Note that this does not acknowledge the \gls{ue} if the overall frame has been correctly decoded. A way to do \gls{dlc} frame acknowledgment for this setting is discussed in Section~\ref{sec:practical:ack}.} For simplicity, we assume that if a \glspl{ue} is not capable of sending the entire frame during an access period, it simply drops the current frame. The latter assumption is made to simplify the comparison between the proposed protocol and the legacy ones. The access period is given as $T_{\rm ac} = N_{\rm ac} (L T_{\rm symb} + T_{\rm sw})$, where $L$ is the number of symbols that comprises each packet.

Now assume that each \gls{ue} has its own model $\hat{f}_k$ obtained during the channel oracle module. Different from legacy access protocols -- \emph{e.g.} \gls{saloha} --~\cite{Bertsekas1996,Popovski2020}, the new key idea here is that the \gls{ap} can leverage the channel control provided by the \gls{ris} to spatially coordinate the access of the \glspl{ue} over the reflected angular space. Motivated by 5G New Radio beam sweeping of the synchronization signal block~\cite{3gpp:rel15}, the \gls{ap} can thus control the \gls{ris} to sweep through a new set of configurations, while each active \gls{ue} use its model $\hat{f}_k$ to evaluate and decide if a current configuration related to its corresponding slot is worth to attempt transmission of its packet. This decision process is defined by \emph{access policies}, which can be defined in different ways depending on how \glspl{ue} explore their knowledge $\hat{f}_k$ and the performance metric to be optimized.

For simplicity, we will assume that the number of configurations is equal to the number of access slots. Therefore, the \gls{ap} designs an \emph{access codebook} being denoted as $\Phi_{\rm ac}\equiv\Theta_{\rm ac}$ with index set $\mc{N}_{\rm ac}$ and cardinality $|\mc{N}_{\rm ac}| = N_{\rm ac}$. The design of the access codebook takes into account the range of the angular position of the \glspl{ue}, $\theta_k\in[0,\pi/2]$, and it is designed in such a way that it ensures certain average \gls{snr} requirements to the \glspl{ue} so as to improve the probability of successful transmission. We consider that this access codebook is designed at the deployment and its knowledge is shared with the \glspl{ue}. Fig.~\ref{fig:system-model:channel-oracle} also shows the selection of a single access slot per \gls{ue} after the model $\hat{f}_k$ has been obtained.
Note that the \gls{ap} and \glspl{ue} can still further exploit legacy strategies to improve the probability of successful transmissions, such as the capture effect and packet repetition~\cite{Bertsekas1996,Popovski2020}.

\subsection{Performance Metrics}
In this part, we introduce performance metrics used to evaluate and discuss the protocol in detail. Denote as $\mathcal{K}_{\rm ac}\subseteq\mathcal{K}_{a}$ the set of \glspl{ue} that had their data successfully transmitted after the access period. Then, the \textit{expected probability of access} is $\bar{P}_{\rm ac} =\mathbb{E}\{ \mathrm{Pr}\{k\in \mathcal{K}_{\rm ac}| \forall k\in \mc{K}_a \} \}$, where the expectation is taken w.r.t. noise realizations, \glspl{ue}' positions, and access periods given a fixed $K_a$ value. The \textit{expected overall throughput} and the \textit{expected overall goodput} are then
\begin{align}
    \overline{\mathrm{TP}} &= \mathbb{E} \left\{  K_{\rm ac} \right\} \dfrac{ T_{\rm F}}{ T_{\rm ac} } = \mathbb{E} \left\{  K_{\rm ac} \right\} \dfrac{N_{\rm F} T_{s}}{N_{\rm ac} T_s + N_{\rm ac} T_{\rm sw}} \text{, and}\nonumber\\ \overline{\mathrm{GP}} &= \dfrac{ \mathbb{E} \left\{  K_{\rm ac} \right\} T_{\rm F}}{ T_{\rm ac} +\alpha T_{\rm co}}, \text{ respectively,}
    \label{eq:goodput}
\end{align}
where the expectations are taken as before, and \emph{$\alpha\in\mathbb{R}_{+}$ is a parameter that penalizes the channel oracle overhead depending on how many access periods can be done with a realization of the channel oracle.} The channel oracle overhead is further discussed in Section \ref{sec:practical:overhead}. For comparison, the throughput would be equal to the goodput in the case of \gls{saloha} due to the non-existence of the channel oracle module, but clearly, with a different value of $\mathbb{E} \left\{  K_{\rm ac} \right\}$. Moreover, the above metrics and related ones are highly dependent on the design of the configuration codebook and the access policies, as we shall see in Sect.~\ref{sec:results}.
\section{The Channel Oracle Module}\label{sec:channel-oracle}
We start by designing the channel oracle configuration codebook based on a signal processing interpretation of the \gls{dl} channel coefficient defined in~\eqref{eq:system-model:dl-channel-gain}. Then, we detail the pilot signals received by the \glspl{ue} and discuss a way to determine the number of channel uses $L_{\rm co}$ so as to combat noise. Finally, we show how each \gls{ue} learns its own model of the \gls{dl} channel coefficient function based on interpolation methods.

\subsection{Channel Oracle Configuration Codebook}\label{sec:training:configuration-codebook}
Without loss of generality, consider a \gls{ue} $k$ with $k\in\mathcal{K}$. Its \gls{dl} channel coefficient in \eqref{eq:system-model:dl-channel-gain} can be interpreted as a complex-valued, multidimensional signal continuously varying over time, $t\in(-\infty,+\infty)$, due to wireless transmission, and over the reflected angular space, $\theta\DL_r(t)\in[0,{\pi}/{2}]$, or simply space. Accordingly, we rewrite \eqref{eq:system-model:dl-channel-gain} as:~  $\zeta_{k}\DL(t,\theta\DL_r)={\sqrt{\beta_{k}\DL(t)}}{e^{j \omega \psi_{k}(t)}}{\mathrm{A}_{k}(t,\theta\DL_r(t))}$.
For convenience, we abstract away the time domain by using Assumption \ref{assu:slotted-config}, which states that configurations -- consequently, reflection angles -- can only change once per slot. Thus, if the time domain can be discretized over infinitely many slots, the time domain can be folded onto the space domain without losing continuity. For the sake of the analysis, the signal can then be expressed as:
\begin{equation}
    \zeta\DL_{k}(\theta\DL_r)=\sqrt{\beta\DL_{k}}e^{j\omega\psi_{k}}M_z\sum_{m\in\mathcal{M}_x}e^{j\omega d_x (m+1)(\sin\theta_k-\sin\theta\DL_r)}.
    \label{eq:training:codebook:signal-expression}
\end{equation}
By substituting $\omega=\frac{2\pi}{\lambda}$, $\beta\DL_{k}$ and $\psi_{k}$ from \eqref{eq:system-model:dl-channel-gain} and \eqref{eq:system-model:array-factor}, we get eq.~\eqref{eq:training:codebook:complete-signal-expression} at the top of the next page.
\begin{figure*}[t]
    \begin{align}
        \zeta\DL_{k}(\theta\DL_r)=
        \underbrace{\vphantom{\sum_{m\in\mc{M}_x}} \left(\sqrt{\beta\DL_{k}}M_z\right)}_{\text{Term 1}}
        \underbrace{\vphantom{\sum_{m\in\mc{M}_x}} e^{j2\pi F_0\left(\frac{d_a+d_k}{d_x} - \frac{M_x+1}{2}(\sin\theta_a-\sin\theta_k) \right)}}_{\text{Term 2}}
        \underbrace{\sum_{m\in\mc{M}_x}e^{j2\pi F_0 (m+1)(\sin\theta_k - \sin\theta\DL_r)}}_{\text{Term 3},\,a_{k}(\theta\DL_r)}
        \label{eq:training:codebook:complete-signal-expression}
    \end{align}
    \begin{align}
        c_k(i)=\dfrac{1}{T_p}\int_{0}^{T_p}\hspace{-0.2cm}a_k(\theta\DL_r)e^{-j2\pi F_0 i \theta\DL_r} d\theta\DL_r = \dfrac{1}{T_p}\hspace{-0.1cm}\sum_{m\in\mc{M}_x} \hspace{-0.1cm} e^{j2\pi F_0(m+1)\sin\theta_k}\int_{0}^{\frac{\pi}{2}}\hspace{-0.3cm}e^{-j2\pi F_0 ((m+1) \sin\theta\DL_r + i\theta\DL_r)}d\theta\DL_r
        \label{eq:training:codebook:fourier-coefficient}
    \end{align}
    \begin{align}
        P_{a_k}=\dfrac{1}{T_p}\int_{0}^{T_p}\lvert a_k(\theta\DL_r)\rvert^2d\theta\DL_r\stackrel{(a)}{\leq}\dfrac{1}{T_p}\int_{0}^{\frac{\pi}{2}} \left(\sum_{m\in\mc{M}_x} \left\lvert e^{j2\pi F_0(m+1)(\sin\theta_k-\sin\theta\DL_r)}\right\rvert \right)^2 d\theta\DL_r= M^2_x
        \label{eq:training:codebook:avg-power}
    \end{align}
    \hrule
\end{figure*}
Clearly, this signal is periodic over space with \emph{fundamental spatial frequency} and \emph{fundamental spatial period} respectively given by:~$F_0=\frac{d_x}{\lambda} \quad\text{and}\quad T_p=\frac{\lambda}{d_x}$. Note that typically $d_x=o(\lambda)$, consequently $F_0\leq 1$; meaning that we are dealing with a very slowly varying signal over space. Also, the above signal is \emph{random} due to the unknown position of the \gls{ue} given by the random variables $d_k$ and $\theta_k$.

\noindent \textbf{Codebook Design.}
We can now define the channel oracle configuration codebook $\Phi_{\rm co}$ based on the Nyquist-Shannon theorem.\footnote{In \gls{ris} literature, it is worth mentioning that Nyquist-Shannon theorem was also used in connection with accurate near field channel modeling~\cite{Pizzo2022,Pizzo2022b}, which is totally different from the way we have applied.} The key idea is that the \gls{ap} can design the set of configurations $\Theta_{\rm co}$, equivalently $\Phi_{\rm co}$, by ensuring that $\zeta\DL_{k}(\theta\DL_r)$ is sampled according to the Nyquist-Shannon theorem~\cite{Proakis2006} and taking into account the statistics of the \glspl{ue}' positions $d_k$ and angles $\theta_k$. In such a manner, each \gls{ue} can locally reconstruct its analog signal $\zeta\DL_{k}(\theta\DL_r)$ based on the sampling theory~\cite{Proakis2006,Eldar2015}. Let $T_{\rm samp}\in\mathbb{R}_+$ be the \emph{spatial sampling period} and $F_{\rm samp}=\frac{1}{T_{\rm samp}}$ be the \emph{spatial sampling frequency}. Based on the Nyquist-Shannon theorem, the spatial sampling frequency should satisfy $F_{\rm samp}\geq 2 F_{\max}$, where $F_{\max}\in\mathbb{R}_{+}$ is the \emph{maximum spatial frequency} of the signal $\zeta\DL_{k}(\theta\DL_r)$. \emph{For now, note that $F_{\max}$ depends on $k$, and, consequently does the choice of $F_{\rm samp}$, where we drop the dependency for convenience.} Let then $N_{\rm co}\in\mathbb{Z}_{+}$ denote the number of channel oracle configurations (samples), which is enumerated by $\mathcal{N}_{\rm co}$. Based on the boundaries of $\theta\DL_r\in[0,{\pi}/{2}]$, this number should satisfy the following inequality to ensure perfect reconstruction of the signal $\zeta_k\DL(\theta\DL_r)$ at the $k$-th UE under a noiseless condition:
\begin{equation}
    N_{\rm co}\geq\left\lceil{\frac{\pi}{2}}{{F}_{\rm{ samp}}}\right\rceil.
    \label{eq:channel-oracle:nyquist-number-of-configs}
\end{equation}
By selecting $N_{\rm co}$ accordingly, the channel oracle configuration codebook design is thus:
\begin{equation}
    \Theta_{\rm co}=\{\theta\DL_r[n]:nT_{\rm samp}, n\in\mc{N}_{\rm co}\}\text{ with }\Phi_{\rm co}\stackrel{(a)}{=}h^{-1}(\Theta_{\rm co}),
    \label{eq:training:codebook}
\end{equation}
where $h:\Theta\mapsto\Phi$ following \eqref{eq:system-model:optimal-config}. By using such a codebook, the analog \gls{dl} channel coefficient signal $\zeta\DL_{k}(\theta\DL_r)$ is uniformly sampled over space as
\begin{equation}
    \zeta\DL_{k}[n]=\zeta\DL_{k}(nT_{\rm samp}),
\end{equation}
where $\zeta\DL_{k}[n]$ denotes the discrete-time signal with $n\in\mc{N}_{\rm co}$. To implement such a design, we need to specify $F_{\max}$ and also make it independent on $k$. For this purpose, we analyze the spatial frequency of Term 3 in \eqref{eq:training:codebook:complete-signal-expression}, namely $a_{k}(\theta\DL_r):[0,\pi/2]\mapsto\mathbb{C}$, since the other two terms are independent of $\theta\DL_r$. Thus, $F_{\max}$ is also the maximum spatial frequency of $a_{k}(\theta\DL_r)$. However, the form of $a_k(\theta\DL_r)$ does not allow for an analytical treatment and, hence, we resorted to two approximation methods. 

\noindent \textbf{Approximation 1: Taylor series.} By using the first term of the Taylor series expansion $\sin x=x+\mc{O}(x^3)$, we approximate $a_{k}(\theta\DL_r)$ as:~$\tilde{a}_{k}(\theta\DL_r)\approx\sum_{m\in\mc{M}_x}e^{j2\pi F_0 (m+1) (\sin\theta_k - \theta\DL_r)}.$ The above signal can be seen as a set of harmonically-related complex exponentials~\cite{Proakis2006}, whose highest spatial frequency is associated with the $M_x$-th complex exponential. Then, $F_{\max}$ is approximated by
\begin{equation}
    \tilde{F}_{\max}= M_x F_0 = M_x \dfrac{d_x}{\lambda}= \dfrac{D_x}{\lambda},
    \label{eq:training:codebook:max-freq-1}
\end{equation}
where recall that $D_x$ is the horizontal dimension of the RIS and $(\tilde{\cdot})$ denotes the approximation. Note that this approximation is only good for very small values of $\theta\DL_r$ and already does not depend on $k$.

\noindent \textbf{Approximation 2: Power Conservation.}
Since $a_{k}(\theta\DL_r)$ is periodic, another form to represent it would be to obtain its Fourier series. To do so, we first rewrite $a_k(\theta\DL_r)$ in \eqref{eq:training:codebook:signal-expression} as
\begin{equation}
    a_k(\theta\DL_r)=\left(\sum_{m\in\mc{M}_x}e^{j2\pi F_0 (m+1)(\sin\theta_k - \sin\theta\DL_r)}\right) u(\theta\DL_r),
\end{equation}
where $u(\theta\DL_r)$ is the rectangular function with $u(\theta\DL_r)=1$ if $\theta\DL_r\in[0,\pi/2]$ and $0$ otherwise. First, we remark that the Fourier series of this signal exists because it satisfies the weak Dirichlet conditions~\cite{Proakis2006}, having finite energy in one period. Thus, the Fourier series of $a_{k}(\theta\DL_r)$ can be written as \cite{Proakis2006}:~$a_k(\theta\DL_r)=\sum_{i=-\infty}^{\infty} c_k(i) e^{j2\pi i F_0\theta\DL_r},$ whose coefficient $c_k(i)\in\mathbb{C}$ is calculated as in eq.~\eqref{eq:training:codebook:fourier-coefficient} at the top of this page. Unfortunately, the integral in eq.~\eqref{eq:training:codebook:fourier-coefficient} does not admit a closed-form solution and we had to resort to numerical solutions. With the Fourier series of the signal of interest in hand, we can now evaluate its \emph{average power} $P_{a_k}\in\mathbb{R}_+$ as in eq.~\eqref{eq:training:codebook:avg-power}~\cite{Proakis2006}. In eq.~\eqref{eq:training:codebook:avg-power}, we used the subadditivity property of the absolute value in (a). As expected, eq.~\eqref{eq:training:codebook:avg-power} gives us an upper bound for the power contribution from Term 3 in eq. \eqref{eq:training:codebook:signal-expression}, which meets the expected array gain of $M_x$ coming from the \gls{ris} elements along the $x$-dimension. Now, based on average power conservation, we propose a heuristic approach to approximate the maximum spatial frequency of the signal. First, by the Parseval's relation~\cite{Proakis2006} and the above results, we have that
\begin{equation}
    P_{a_k}=\sum_{i=-\infty}^{\infty}|c_k(i)|^2\leq M^2_x.
    \label{eq:training:codebook:avg-power-bound}
\end{equation}
Let $0\leq\epsilon\leq1$ be an \emph{efficiency parameter} that parameterizes the notion of conservation efficiency of the average power $P_{a_k}$, \emph{i.e.}, it measures the percentage of the error we will commit due to the approximation. Then, the smallest symmetric interval of the coefficients of the series that ensures a desired power efficiency $\epsilon$ is given by:
\begin{equation}
    \text{find } {I^{\epsilon}_k\in\mathbb{Z}_{+}} \text{ s.t. } \sum_{i=-I^{\epsilon}_k}^{I^{\epsilon}_k}|c_k(i)|^2 \geq (1-\epsilon) P_{a_k},
\end{equation}
where the existence of a solution is ensured by the fact that the infinite sum of coefficients is bounded in \eqref{eq:training:codebook:avg-power-bound}, $\exists I^{\epsilon}_k \in\mathbb{Z}_{+},\,\forall k$. Thus, the maximum spatial frequency $F_{\max}$ can be approximated as:
\begin{equation}
    \tilde{F}^{\epsilon}_{\max}=I^{\epsilon}_k \cdot F_0,
    \label{eq:training:codebook:max-freq-2}
\end{equation}
where $F_{\max}\geq \tilde{F}^{\epsilon}_{\max}$ with equality when $\epsilon\xrightarrow[]{}0$. In practice, $F_0$ is fixed, while $I^\epsilon_k$ depends on the \gls{ue}'s position through $\theta_k$. Hence, note that different from Approximation 1, Approximation 2 depends on $\epsilon$ and the position of the \gls{ue}.

\noindent \textbf{Evaluating Approximations.}
The two approximation methods proposed are evaluated in Fig.~\ref{fig:training:codebook:fmax} as a function of the \gls{ue}'s angles and 
a spatial fundamental frequency of $F_0=0.5$. For Approximation 1, the approximated maximum (spatial) frequency evaluates as $\tilde{F}_{\max}=M_xF_0=50$, $\forall k$. While for Approximation 2, we evaluated different values of $\epsilon$ as $10^{-1}$, $10^{-2}$, and $10^{-3}$. The figure shows that the maximum frequency is highly dependent on the position of the \glspl{ue}. This is undesired from the standpoint of the \gls{ap} and the codebook design since it is unaware of the \glspl{ue} positions in advance. For this reason, we evaluate the following statistics of $\tilde{F}^{\epsilon}_{\max}$ over $\theta_k$ that are going to be relevant in the sequel. The medians w.r.t. $\theta_k$ are: $3.0$, $5.0$, and $14.5$ for $\epsilon$ equals to $10^{-1}$, $10^{-2}$, and $10^{-3}$. On the other hand, the maximums w.r.t. the angle $\theta_k$ are: $6.5$, $45.0$, and $186.0$. 
In general, we note that Approximation 2 is a more accurate method and that the smaller the $\epsilon$, the better the characterization of the maximum frequency. This consequently means that the signal will be better discretized according to the Nyquist-Shannon theorem~\cite{Proakis2006}, allowing a better reconstruction of the signal of interest at the \gls{ue}'s side.

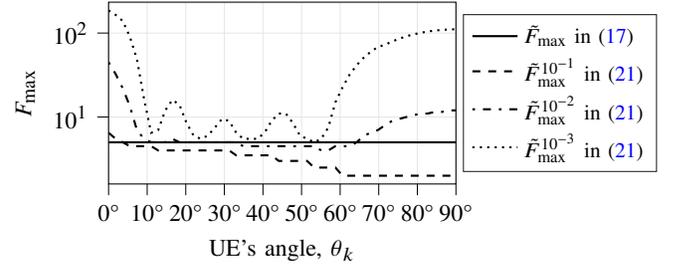
\begin{figure}[t]
    \centering
\begin{tikzpicture}

\definecolor{lavender233}{RGB}{233,233,233}
\definecolor{lightgray204}{RGB}{204,204,204}

\begin{axis}[
width=0.7\columnwidth,
height=4cm,
legend cell align={left},
legend style={
  at={(1.02,0.5)},
  anchor=west,
},
log basis y={10},
tick align=outside,
tick pos=left,
xlabel={UE's angle, \(\displaystyle \theta_k\)},
xmajorgrids,
xmin=0, xmax=90,
xtick={0,10,20,30,40,50,60,70,80,90},
xticklabels={0°,10°,20°,30°,40°,50°,60°,70°,80°,90°},
ylabel={
$F_{\max}$},
ymajorgrids,
ymin=1.5944314354777, ymax=233.312008106856,
ymode=log,
ytick={0.1,1,10,100,1000,10000},
yticklabels={
  \(\displaystyle {10^{-1}}\),
  \(\displaystyle {10^{0}}\),
  \(\displaystyle {10^{1}}\),
  \(\displaystyle {10^{2}}\),
  \(\displaystyle {10^{3}}\),
  \(\displaystyle {10^{4}}\)
},
tick label style={font=\small},
label style={font=\small},
]
\addplot [thick, black]
table {%
0 5
1.83673469387755 5
3.6734693877551 5
5.51020408163265 5
7.3469387755102 5
9.18367346938775 5
11.0204081632653 5
12.8571428571429 5
14.6938775510204 5
16.530612244898 5
18.3673469387755 5
20.2040816326531 5
22.0408163265306 5
23.8775510204082 5
25.7142857142857 5
27.5510204081633 5
29.3877551020408 5
31.2244897959184 5
33.0612244897959 5
34.8979591836735 5
36.734693877551 5
38.5714285714286 5
40.4081632653061 5
42.2448979591837 5
44.0816326530612 5
45.9183673469388 5
47.7551020408163 5
49.5918367346939 5
51.4285714285714 5
53.265306122449 5
55.1020408163265 5
56.9387755102041 5
58.7755102040816 5
60.6122448979592 5
62.4489795918367 5
64.2857142857143 5
66.1224489795918 5
67.9591836734694 5
69.7959183673469 5
71.6326530612245 5
73.469387755102 5
75.3061224489796 5
77.1428571428571 5
78.9795918367347 5
80.8163265306122 5
82.6530612244898 5
84.4897959183673 5
86.3265306122449 5
88.1632653061224 5
90 5
};
\addlegendentry{$\tilde{F}_{\max}$ in~\eqref{eq:training:codebook:max-freq-1}}
\addplot [thick, black, dashed]
table {%
0 6.5
1.83673469387755 5.5
3.6734693877551 5
5.51020408163265 4.5
7.3469387755102 4.5
9.18367346938775 4.5
11.0204081632653 4.5
12.8571428571429 4
14.6938775510204 4
16.530612244898 4
18.3673469387755 4
20.2040816326531 4
22.0408163265306 4
23.8775510204082 4
25.7142857142857 4
27.5510204081633 4
29.3877551020408 4
31.2244897959184 4
33.0612244897959 3.5
34.8979591836735 3.5
36.734693877551 3.5
38.5714285714286 3.5
40.4081632653061 3.5
42.2448979591837 3.5
44.0816326530612 3
45.9183673469388 3
47.7551020408163 3
49.5918367346939 3
51.4285714285714 3
53.265306122449 2.5
55.1020408163265 2.5
56.9387755102041 2.5
58.7755102040816 2.5
60.6122448979592 2
62.4489795918367 2
64.2857142857143 2
66.1224489795918 2
67.9591836734694 2
69.7959183673469 2
71.6326530612245 2
73.469387755102 2
75.3061224489796 2
77.1428571428571 2
78.9795918367347 2
80.8163265306122 2
82.6530612244898 2
84.4897959183673 2
86.3265306122449 2
88.1632653061224 2
90 2
};
\addlegendentry{$\tilde{F}^{10^{-1}}_{\max}$ in~\eqref{eq:training:codebook:max-freq-2}}
\addplot [thick, black, dash pattern=on 1pt off 3pt on 3pt off 3pt]
table {%
0 45
1.83673469387755 32.5
3.6734693877551 22
5.51020408163265 13.5
7.3469387755102 7
9.18367346938775 5.5
11.0204081632653 5
12.8571428571429 5
14.6938775510204 5
16.530612244898 5.5
18.3673469387755 5
20.2040816326531 5
22.0408163265306 5
23.8775510204082 5
25.7142857142857 5
27.5510204081633 5
29.3877551020408 5
31.2244897959184 5
33.0612244897959 5
34.8979591836735 4.5
36.734693877551 4.5
38.5714285714286 4.5
40.4081632653061 4.5
42.2448979591837 4.5
44.0816326530612 4.5
45.9183673469388 4.5
47.7551020408163 4.5
49.5918367346939 4.5
51.4285714285714 4.5
53.265306122449 4.5
55.1020408163265 4
56.9387755102041 4
58.7755102040816 4.5
60.6122448979592 4.5
62.4489795918367 4.5
64.2857142857143 5
66.1224489795918 6
67.9591836734694 6.5
69.7959183673469 7
71.6326530612245 8
73.469387755102 9
75.3061224489796 9.5
77.1428571428571 10
78.9795918367347 10.5
80.8163265306122 11
82.6530612244898 11
84.4897959183673 11.5
86.3265306122449 11.5
88.1632653061224 12
90 12
};
\addlegendentry{$\tilde{F}^{10^{-2}}_{\max}$ in~\eqref{eq:training:codebook:max-freq-2}}
\addplot [thick, black, dotted]
table {%
0 186
1.83673469387755 164.5
3.6734693877551 139.5
5.51020408163265 90.5
7.3469387755102 49
9.18367346938775 16.5
11.0204081632653 6.5
12.8571428571429 7
14.6938775510204 12
16.530612244898 16
18.3673469387755 14
20.2040816326531 8.5
22.0408163265306 6
23.8775510204082 5.5
25.7142857142857 6
27.5510204081633 7.5
29.3877551020408 9.5
31.2244897959184 9
33.0612244897959 6.5
34.8979591836735 5.5
36.734693877551 5.5
38.5714285714286 5.5
40.4081632653061 6.5
42.2448979591837 9
44.0816326530612 11
45.9183673469388 11
47.7551020408163 8.5
49.5918367346939 6
51.4285714285714 5.5
53.265306122449 5
55.1020408163265 5.5
56.9387755102041 7.5
58.7755102040816 15
60.6122448979592 22.5
62.4489795918367 32
64.2857142857143 41.5
66.1224489795918 51
67.9591836734694 60
69.7959183673469 68.5
71.6326530612245 72.5
73.469387755102 79.5
75.3061224489796 85.5
77.1428571428571 91.5
78.9795918367347 96.5
80.8163265306122 100.5
82.6530612244898 104.5
84.4897959183673 107
86.3265306122449 109
88.1632653061224 110.5
90 111
};
\addlegendentry{$\tilde{F}^{10^{-3}}_{\max}$ in~\eqref{eq:training:codebook:max-freq-2}}
\end{axis}
\end{tikzpicture}
    \vspace{-6mm}
    \caption{\small Evaluation of the two approximations of the maximum spatial frequency $F_{\max}$ of $a_k(\theta\DL_r)$ and $\zeta\DL_k(\theta\DL_r)$ as a function of the UE's angle $\theta_k$ with $F_0=0.5$ and different values of $\epsilon$.}
    \label{fig:training:codebook:fmax}
\end{figure}

\noindent \textbf{Approximating $N_{\rm co}$.} We now approximate \eqref{eq:channel-oracle:nyquist-number-of-configs} based on the the approximation of $F_{\max}$. Based on Approximation 2 in~\eqref{eq:training:codebook:max-freq-2}, the $F_{\rm samp}$ that allows perfect reconstruction of the signal is~\cite{Proakis2006}:
\begin{align}
    F_{\rm samp}\geq 2 F_{\max} \implies 
    \tilde{F}^{\epsilon}_{\mathrm{samp}} \gtrapprox 2\tilde{F}^{\epsilon}_{\max}, \label{eq:training:codebook:nyquist-shannon} \text{ and }\\
    \tilde{F}^{\epsilon}_{\max}=I^{\epsilon}_k F_0=I^{\epsilon}_k\dfrac{d_x}{\lambda}=\dfrac{I^{\epsilon}_k}{\lambda M_x}{D_x},
\end{align}
where $\gtrapprox$ denotes an approximation of the inequality. The above relationship shows that the approximated spatial sampling frequency $\tilde{F}^{\epsilon}_{\mathrm{samp}}\in\mathbb{R}_{+}$ is directly proportional to the horizontal size of the \gls{ris}, $D_x$, and depends on both the conservation efficiency, $\epsilon$, and the UE's position, $\theta_k$. Moreover, it is important to note that we will always perform some undersampling because of $F_{\max}>\tilde{F}^{\epsilon}_{\max}$ since $\epsilon$ cannot be made infinitely small. Similar conclusions can be obtained when considering Approximation 1 in \eqref{eq:training:codebook:max-freq-1}. From \eqref{eq:training:codebook:nyquist-shannon}, we obtain an approximate lower bound on the number of configurations (samples) as:
    $N_{\rm co}\geq\left\lceil{\frac{\pi}{2}}{\tilde{F}^{\epsilon}_{\rm{ samp}}}\right\rceil=\left\lceil \pi \tilde{F}^{\epsilon}_{\max}\right\rceil.$
The remaining problem with using this approximated result to design the codebook is the dependence of the choice of $N_{\rm co}$ on the \glspl{ue}' positions, which are unknown to the \gls{ap}. Thus, we consider three different statistical criteria to get feasible choices of $N_{\rm co}$ that are independent of the \glspl{ue}' position and would heuristically and statistically ensure good, but different, reconstruction performances: \textit{Median:} choose ${N}_{\rm co}=\lceil \pi \mathrm{med}_k\{\tilde{F}^{\epsilon}_{\max,k}\}\rceil$, meaning that half of the \glspl{ue} will statistically have their approximated lower respected; \textit{Maximum:} choose ${N}_{\rm co}=\lceil \pi \mathrm{max}_k\{\tilde{F}^{\epsilon}_{\max,k}\}\rceil$, meaning that all the \glspl{ue} will have their approximated lower bound respected, but at the price of increased duration of the training phase due to oversampling for most of the \glspl{ue}; \textit{Taylor-approximation:} choose ${N}_{\rm co}$ according to \eqref{eq:training:codebook:max-freq-1}, resulting in some oversampling.

\begin{remark}\label{remark:n-co-values}
    Following the results from Fig. \ref{fig:training:codebook:fmax} by assuming $F_0=0.5$, we have ${N}_{\rm co}=16$ (median) and ${N}_{\rm co}=142$ (maximum) for $\epsilon=10^{-2}$; ${N}_{\rm co}=46$ (median) and ${N}_{\rm co}=585$ (maximum) for $\epsilon=10^{-3}$; and ${N}_{\rm co}=150$ configurations for Approximation~1. 
\end{remark}

\subsection{Pilot Signals}\label{sec:training:training-signal}
For \glspl{ue} to get input data, the \gls{ap} transmits pilot signals toward the \glspl{ue}, while controlling the \gls{ris} to sweep through the configuration codebook designed according to~\eqref{eq:training:codebook}. For $n\in\mc{N}_{\rm co}$ and $k\in\mathcal{K}$, the \gls{dl} pilot signal $\mathbf{w}_{k}[n]\in\mathbb{C}^{L_{\rm co}}$ received by the $k$-th UE is:
    $\mathbf{w}_{k}[n]=\sqrt{\rho_a}\zeta\DL_{k}[n]\boldsymbol{\upsilon}_{\rm co} + \boldsymbol{\eta}_{k}[n],$
where $\rho_a$ is the \gls{ap} transmit power, $\boldsymbol{\upsilon}_{\rm co}\in\mathbb{C}^{L_{\rm co}}$ denotes the pilot symbol with zero mean and $\mathbb{E}\{\ltwonorm{\boldsymbol{\upsilon}_{\rm co}}^2\}=L_{\rm co}$, and $\boldsymbol{\eta}_k[n]\in\mathbb{C}^{L_{\rm co}}\sim\mc{N}_\mathbb{C}(\mathbf{0},\sigma^2\mathbf{I}_{L_{\rm co}})$ is the receiver noise with variance~$\sigma^2$. We assume that noise is \gls{iid} over $n$. The final goal of the $k$-th UE is to reconstruct the analog signal $\zeta\DL_{k}(\theta\DL_r)$ in \eqref{eq:training:codebook:signal-expression} from the collection of samples $\{\mathbf{w}_{k}[n]\}_{n\in\mc{N}_{\rm co}}$. Before doing so, the $k$-th UE combats the receiver noise by estimating the sampled complex amplitudes $\zeta\DL_{k}[n]$ from $\mathbf{w}_{k}[n]$, whose process is summarized in the following corollary.

\begin{corollary}
    The \gls{crlb} for the estimation of $\zeta\DL_{k}[n]$ from $\mathbf{w}_{k}[n]$ is  $\delta\DL_{\rm tol}\geq \frac{1}{\mathrm{SNR}\DL_a L_{\rm co}},$ where $\mathrm{SNR}\DL_a=\frac{\rho_a}{\sigma^2}$ is the \gls{dl} transmit \gls{snr} and $\delta\DL_{\rm tol}$ is a chosen estimation error tolerance. From the \gls{crlb}, we obtain the minimum variance unbiased estimator as: 
    \begin{equation}
        \hat{\zeta}\DL_{k}[n]=\dfrac{1}{L_{\rm co}\sqrt{\rho_a}}\boldsymbol{\upsilon}_{\rm co}^\transp\mathbf{w}_k[n], \text{ and } \mc{N}_{\mathbb{C}}(\zeta\DL_{k}[n],\delta\DL_{\rm tol}).
        \label{eq:training:dl-received:mvu}
    \end{equation}
    \label{cor:training:estimation}  
\end{corollary}
\begin{proof}
    The proof follows the steps in \cite[Ch. 13]{Kay1997}.
\end{proof}
\noindent Consequently, the number of channel uses can be chosen as 
\begin{equation}
    L_{\rm co}\geq\left\lceil\dfrac{1}{\mathrm{SNR}\DL_a\delta\DL_{\rm tol}}\right\rceil.
    \label{eq:training:dl-received:channel-uses-bound}
\end{equation}

\subsection{Learning Channel Model}\label{sec:training:information}
Now that the \glspl{ue} got their estimates $\{\hat{\zeta}\DL_{k}[n]\}_{n\in\mc{N}_{\rm co}}$ following Corollary \ref{cor:training:estimation}, they can obtain their own model $\hat{f}_{k}$ such that $\hat{\zeta}\DL_k=\hat{f}_k(\theta\DL_r)$. Let $\hat{\zeta}\DL_{k}(\theta\DL_r)$ denote the signal resulting from an interpolation process over the collection of estimates $\{\hat{\zeta}\DL_{k}[n]\}_{n\in\mc{N}_{\rm co}}$. Let $\Lambda:[0,\pi/2]\mapsto\mathbb{C}$ be an interpolating function. Then, the reconstructed signal can be written as
\begin{equation}
    \hat{\zeta}\DL_{k}(\theta\DL_r)=\sum_{n\in\mc{N}_{\rm co}} \hat{\zeta}\DL_{k}[n]\cdot \Lambda(\theta\DL_r-nT_{\rm samp}),\,\forall \theta\DL_r \in \Theta_{\rm co}.
    \label{eq:training:info:reconstruction}
\end{equation}
The above reconstruction can give the parameters necessary to define a model $\hat{f}_k$ by using interpolation theory \cite{Hamming1986}. In the following corollary, we characterize the expected \gls{se} of the reconstruction w.r.t. $\{\hat{\zeta}\DL_{k}[n]\}_{n\in\mc{N}_{\rm co}}$.
\begin{corollary}\label{cor:training:reconstruction}   
   The interpolation result is distributed as
        $\hat{\zeta}\DL_{k}(\theta\DL_r)\sim\mc{N}_{\mathbb{C}}( \mathring{\zeta}\DL_{k}(\theta\DL_r),\delta\DL_{\rm tol}\sum_{n\in\mc{N}_{\rm co}}\Lambda(\theta\DL_r-nT_{\rm samp})),$
   where $\mathring{\zeta}\DL_{k}(\theta\DL_r)=\sum_{n\in\mc{N}_{\rm co}} {\zeta}\DL_{k}[n]\cdot \Lambda(\theta\DL_r-nT_{\rm samp})$ denotes the interpolation result under a noiseless condition. Thus, the expected \gls{se} $\mathbb{E}\{|\hat{\zeta}\DL_{k}(\theta\DL_r)-{\zeta}\DL_{k}(\theta\DL_r)|^2\}$ can be computed as
    \begin{equation} 
        \overline{\mathrm{SE}}=\delta\DL_{\rm tol}\sum_{n\in\mc{N}_{\rm co}}\Lambda(\theta\DL_r-nT_{\rm samp}) + \mathrm{TSE},
        \label{eq:training:expectedSE}
    \end{equation}
    where the first term on the right-hand side accounts for the noise and estimation effects, while the second, namely $\mathrm{TSE}$, is the true \gls{se}, referring just to the error incurred by the interpolation and sampling processes.
\end{corollary}
\begin{proof}
    The proof straightforwardly follows from Corollary \ref{cor:training:estimation} and eq.~\eqref{eq:training:info:reconstruction}.
\end{proof}
This corollary indicates that the reconstruction performance depends on i) the estimation performance of $\{\hat{\zeta}\DL_{k}[n]\}_{n\in\mc{N}_{\rm co}}$, which is based on the choice of $\delta\DL_{\rm tol}$, and ii) the performance of the interpolation process, which is measured by the $\mathrm{TSE}$. In practice, the latter depends on the choice of the interpolation method (\emph{e.g.}, linear, cubic, spline~\cite{Hamming1986}). Therefore, we conclude that there is a clear trade-off between the duration of the channel oracle module and the quality of the model $\hat{f}_k$: the smaller $\delta\DL_{\rm tol}$, the longer the channel oracle ($\uparrow L_{\rm co}$), and, consequently, the better would be the expected reconstruction performance as measured by $\overline{\mathrm{SE}}$.
\section{The Access Module}\label{sec:access}
We start by designing an access configuration codebook, whose design goal is to cover the area of interest while ensuring that the \gls{ul} \gls{snr} is greater than a minimum threshold regardless of the position of the \glspl{ue} so as to improve the probability that the \gls{ap} successfully decodes their packets. Then, we propose different access policies based on the channel models learned by \glspl{ue} and detail the \gls{ul} received signal at the \gls{ap}. 

\subsection{Access Configuration Codebook}\label{sec:access:codebook}
A straightforward design for the access codebook would be to uniformly slice the angular domain $\theta\DL_r\in[0,\pi/2]$ into $N_{\rm ac}$ slices. However, two problems occur when considering this design: a) there is no guarantee on the value of the \gls{ul} received \gls{snr} at the \gls{ap} from the \glspl{ue}; b) the main lobe of the array factor in \eqref{eq:system-model:phase-shifted-array-factor} has a width that depends on the reflection angle $\theta\DL_r$~\cite{Balanis2012antenna}, where the higher the value of $\theta\DL_r$, the wider the main lobe. To solve these drawbacks, we introduce a tailored design for the access codebook $\Phi_{\mathrm{ac}}$ together with a power control strategy that is carried out by the \glspl{ue}. More formally, we want to design an access codebook and the \gls{ue}'s transmit power in order to have at least one configuration $n\in\mc{N}_{\rm ac}$ that satisfies the following for any \gls{ue} $k\in\mc{K}$
\begin{equation}
    \mathrm{SNR}\UL_k \beta_{k}\UL \left\lvert\mathrm{A}_{k}(\theta\DL_r[n])\right\rvert^2 \geq \gamma_{\rm ac},
    \label{eq:access:codebook:minimum-snr-inequality}
\end{equation}
where $\mathrm{SNR}\UL_k=\frac{\rho_k}{\sigma^2}$ is the \gls{ul} transmit \gls{snr} with $\rho_k$ being the \gls{ul} transmit power and $\sigma^2$ being the noise power at the \gls{ap}. The \textit{threshold decoding \gls{snr}} $\gamma_{\rm ac}\in\mathbb{R}_{+}$ depends on the decoding capabilities of the \gls{ap}. Observe that the left-hand term is the received \gls{snr} at the \gls{ap} when just a single \gls{ue} transmits over a channel use (that is, $L=1$) and that only $\mathrm{A}_{k}(\theta\DL_r[n])$ is a function of the reflection angle $\theta\DL_r$. To obtain such access codebook design, we carry out two steps: {step 1} certifies that $\left\lvert\mathrm{A}_{k}(\theta\DL_r[n])\right\rvert^2$ gives a minimum gain, while {step 2} ensures that the \glspl{ue} adjust their transmit powers $\rho_k$ accordingly so as to meet the threshold decoding \gls{snr} $\gamma_{\rm ac}$.

\noindent \textbf{Step 1.} From \eqref{eq:system-model:phase-shifted-array-factor}, the normalized power of the array factor can be rewritten as~\cite{Balanis2012antenna, tang2020wireless}
\begin{equation}
    \frac{|\mathrm{A}_{b,k}(\theta\DL_r[n])|^2}{M^2} = \left|\frac{\sin\left(\frac{ \omega d_x }{2} M_x (\sin\theta_k - \sin\theta\DL_r[n])\right)}{M_x \sin\left(\frac{\omega d_x}{2} (\sin\theta_k - \sin\theta\DL_r[n])\right)} \right|^2.
    \label{eq:access:codebook:phased}
\end{equation}
The above expression is the array factor of a linear phased array, which is a periodic function of the angular position having a main lobe of magnitude 1 (0 dB) centered in $\sin\theta_r\DL[n] = \sin\theta_k$ -- which can be well approximated by the main lobe of a $\sinc$ function -- and a side lobe level (SLL) value of approximately 0.045 (-13.46 dB)~\cite{Balanis2012antenna}. Thus, it is possible to design the access codebook letting the main lobe of two consequent configurations overlap at the angular point which provides the desired minimum gain so that each \gls{ue} can always find at least one suitable configuration, regardless of its position. Let then $\tau\in(0.045,1]$ be the minimum gain desired, where the lower limit is set to unambiguously discriminate the main lobe from the side lobes. We want to set $x$ such that $|\sinc(x)|^2 \ge \tau$. Defining $\pm x_\tau$ as the $x$ that satisfy $|\sinc(x_\tau)|^2 = \tau$, the condition~is:
$
- x_\tau < \pi F_0 M_x \left( \sin\theta_k - \sin\theta\DL_r[n] \right) < x_\tau,
$
where recall that $F_0=\frac{d_x}{\lambda}$. Now, define as $\theta^{\scriptscriptstyle \rm DL, \tau+}_r[n]$ and $\theta^{\scriptscriptstyle \rm DL, \tau-}_r[n]$ the right and left angular directions where the gain of the main lobe is precisely $\tau$, namely $\tau$-angular directions. They can be obtained from the following relations
\begin{align} 
        \pi F_0 M_x \left( \sin\theta^{\scriptscriptstyle \rm DL, \tau+}_r[n] - \sin\theta\DL_r[n] \right) = x_\tau \text{, and} \\
        \pi F_0 M_x \left( \sin\theta^{\scriptscriptstyle \rm DL, \tau-}_r[n] - \sin\theta\DL_r[n] \right) = -x_\tau.
    \label{eq:access:codebook:halpowerbw}
\end{align}
To cover the whole area of interest (see Fig. \ref{fig:system-setup}), we impose that $\theta^{\scriptscriptstyle \rm DL, \tau+}_r[N-1]= \pi / 2$, meaning that the last configuration $n=N-1$ has the left $\tau$-angular direction toward the most left direction of the area of interest. Hence, by using the above relationships, we have
$
\sin\theta\DL_r[N-1] = 1 - \frac{x_\tau}{\pi F_0 M_x}.
$
By applying this to compute $\theta^{\scriptscriptstyle \rm DL, \tau-}_r[N - 1]$, we obtain
$
\sin\theta^{\scriptscriptstyle \rm DL, \tau-}_r[N-1] = 1 - 2\frac{x_\tau}{\pi F_0 M_x} = \theta^{\scriptscriptstyle \rm DL, \tau+}_r[N-2],
$
which is then set to overlap the left $\tau$-angular direction of configuration $N-2$. By iterating the procedure, we get the following
        $\sin\theta\DL_r[n] = 1 - (2(N-n)-1) \frac{x_\tau}{\pi F_0 M_x},\,\forall n\in\mc{N}_{\rm ac}.$ 
    \label{eq:access:codebook:power-slicing-rule}
Then, a lower bound on the number of access configurations needed to cover the whole area while incurring a minimum gain of $\tau$ is
\begin{equation}
        N_{\mathrm{ac}}\geq \min\left\{n \,|\, \sin\theta^{\scriptscriptstyle \rm DL, \tau-}_r[n] < 0,\,n\in\mathbb{Z}_{+}\right\} = \left\lceil {\pi} \left(\dfrac{M_x}{2x_{\tau}}\right) F_0\right\rceil.
    \label{eq:access:lower-bound-access-slots}
\end{equation}
The access configuration codebook $\Phi_{\rm ac}\equiv\Theta_{\rm ac}$ is then constructed based on the iterative method defined above given that $N_{\rm ac}$ is chosen according to the bound. Without loss of generality, we will consider $\tau = 0.5$ ($-3$ dB) based on classical literature, which gives $x_\tau \approx 1.391$~\cite{Balanis2012antenna}. 

\noindent \textbf{Step 2.} Given an access codebook following the design of eqs.~\eqref{eq:access:codebook:power-slicing-rule}-\eqref{eq:access:lower-bound-access-slots}, there is at least one configuration, say $n^*$, providing a received \gls{ul} \gls{snr} for \gls{ue} $k$ at the \gls{ap} of
    $\frac{\rho_k}{\sigma^2} \beta_{k}\UL \left\lvert \mathrm{A}_{k}(\theta\DL_r[n^*])\right\rvert^2 \ge \frac{\rho_k}{\sigma^2} \beta_{k}\UL M^2 \tau > \gamma_{\rm ac},$
where the last inequality is imposed to meet the condition in~\eqref{eq:access:codebook:minimum-snr-inequality}. To satisfy the aforementioned condition, we devise a power control policy based on selecting the minimum \gls{ul} transmit power at the \gls{ue}'s side. Since $\beta_{k}\UL$ is a function of the random position of \gls{ue} $k$, we assure the above inequality in the average sense w.r.t. the \gls{ue}'s position as follows: $\rho_k \ge ({\sigma^2}\gamma_{\rm ac}) / ({\mathbb{E}_k\{\beta_{k}\UL\} M^2 \tau}).$\footnote{For the sake of analysis, we keep the power of all \glspl{ue} the same. Nevertheless, each \gls{ue} could estimate their highest \gls{ul} channel gains to derive a power control policy over different statistics other than average.} If we assume that $d_k$ and $\theta_k$ are independent, the expectation of the pathloss evaluates to
\begin{align} 
    \mathbb{E}_k\{\beta_{k}\UL\}=\dfrac{G_a G_k}{(4 \pi )^2} \left(\dfrac{d_x d_z}{d_a} \right)^2  \dfrac{\log(d_\text{max}) -\log(d_\text{min})}{d_\text{max}^2 - d_\text{min}^2},
    \label{eq:access:codebook:expected-pathloss}
\end{align}
based on the \glspl{pdf}:
\begin{align}
    p_{d_k}(d)=\frac{2 d}{(d^2_{\max}-d^2_{\min})}, \text{ for } d_{\min}\leq d\leq d_{\max}\text{ and }\\
    p_{\theta_k}(\theta)=\frac{2}{\pi}, \text{ for } 0\leq {\theta} \leq \frac{\pi}{2}.
    \label{eq:pdfs}
\end{align}

\subsection{Access Policies}\label{sec:access:access-policy}
Based on $\hat{f}_{k}$ obtained in the channel oracle module, each active \gls{ue} can now locally decide in which access slots to transmit its packets by stipulating and following an \emph{access policy}, $\forall k\in\mathcal{K}_a$. In principle, an access policy would like to satisfy two conditions: i)~maximize the \gls{ul} \gls{snr} received at the \gls{ap} for each \gls{ue} so as to improve its probability of access, and ii) reduce the overall probability of collisions among \glspl{ue}. To satisfy the first condition, since the \gls{ul} received \gls{snr}~\eqref{eq:access:codebook:minimum-snr-inequality} is proportional to the \gls{ul} channel gain $|\zeta\UL_k|^2$, a \gls{ue} would like to transmit a packet during access slots associated to \emph{good} configurations or reflection angle $\theta\UL_r[n]$ in respect to its position. Where \emph{goodness} here means high values of channel gains $|{{\zeta}}\UL_{k}[n]|^2$. Based on channel reciprocity (see eq. \eqref{eq:system-model:optimal-config}), a \gls{ue} can measure the \textit{goodness of the access slots} by getting: ${\Hat{\zeta}}\UL_{k}[n]=\hat{f}^{*}_k(\theta\UL_r[n]),\,\forall n\in\mc{N}_{\rm ac},$ where ${\Hat{\zeta}}\UL_{k}[n]\in\mathbb{C}$ is the inferred \gls{ul} channel coefficient for the $n$-th access slot at the \gls{ue}'s side. We let ${\Hat{\boldsymbol{\zeta}}}\DL_{k}\in\mathbb{C}^{N_{\rm ac}}$ denote the collection of inferences. Thus, the \gls{ue} can exploit ${\Hat{\boldsymbol{\zeta}}}\DL_{k}$ so as to choose to transmit during good access slots. To satisfy the second condition and since the \glspl{ue} cannot coordinate among themselves, we consider the transmission of multiple replicas of a packet.

\noindent\textbf{Formal Definition.} Based on the above discussion, we are now ready to formally define an access policy. Let $\Pi_k\subseteq\mc{N}_{\rm ac}$ denote the access set of the $k$-th \gls{ue}, which contains the access slots in which the $k$-th UE will attempt to send its packets, where $|\Pi_k|=RN_{\rm F}$ and $R\in\mathbb{Z}_{+}$, defines the number of replicas to send per packet. To obtain its $\Pi_k$, a \gls{ue} first quantifies the goodness of the access slots based on an \emph{acquisition function} $q:\mathbb{C}\mapsto\mathbb{R}$ that uses an entry of the inferred information ${\Hat{\boldsymbol{\zeta}}}\DL_{k}$ as an input. With the measured qualities of the access slots in hand, the \gls{ue} applies a \emph{selection function} $s:\mathbb{R}\mapsto\mc{N}_{\rm ac}$, which actually defines how to build the access set. By making a parallel to the reinforcement learning literature~\cite{Bishop2006}, the acquisition function can be learned or specified, while the selection function can be deterministic or stochastic. For simplicity, we heuristically consider $q=|\cdot|$ in this work. We now propose three different access policies.
%
\textbf{1. {\texorpdfstring{$R$}{R}-configuration-aware random policy (\texorpdfstring{$R$}{R}-CARAP)}.} 
The \gls{ue} can compute a probability mass function $\mathbf{p}\in\mathbb{R}^{N_{\rm ac}}$ where the $n$-th element of $\mathbf{p}$ is given as $P_n={q({\Hat{\zeta}}\UL_{k}[n])}/{\sum_{n'=1}^{N_{\rm ac}}q({\Hat{\zeta}}\UL_{k}[n'])}.$ The selection function $s$ is then a random function comprised of sampling without replacement of the elements from the set $\mathcal{N}_{\rm ac}$ based on $\mathbf{p}$, $R$, and $N_{\rm F}$. The construction of $\Pi_k$ finalizes when the specified $RN_{\rm F}$ is reached.
%
\textbf{2. {\texorpdfstring{$R$}{R}-greedy-strongest-configurations access policy (\texorpdfstring{$R$}{R}-GSCAP)}.}
This access policy simply works by getting the best configuration $n^{\star}=\argmax_{n\in\mc{N}_{\rm ac}}q({\Hat{\zeta}}\UL_{k}[n])$ without replacement and successively update the set $\Pi_k=\Pi_k\cup \{n^{\star}\}$ until the specified $RN_{\rm F}$ is reached.
%
\textbf{3. {Strongest-minimum access policy (SMAP)}.}
Different from the others, this is the only policy that is not defined for any number of multiple replicas. SMAP simply follows from a heuristic of transmitting only two replicas, $R=2$, according to the following. The first replica of a packet is transmitted during the best access slot, that is, $n_1=\argmax_{n\in\mc{N}_{\rm ac}}q({\Hat{\zeta}}\UL_{k}[n])$. Whereas the second replica is transmitted into the access slot that is closest to ensure the \gls{ul} minimum \gls{snr} threshold $\gamma_{\rm ac}$, which can be written as $n_2=\argmin_{n\in\mc{N}_{\rm ac}\setminus \{n_1\}} \{ \mathrm{SNR}\UL_k \lvert{{\Hat{\zeta}}\UL_{k}[n]}\rvert^2 - \gamma_{\rm ac} \, | \, \mathrm{SNR}\UL_k \lvert{{\Hat{\zeta}}\UL_{k}[n]}\rvert^2 \geq \gamma_{\rm ac} \}$. Then, $\Pi_k=\{n_1,n_2\}$. If $n_2$ does not exist, the \gls{ue} transmits just in slot $n_1$.

\subsection{Access Transmissions}
During the access module, the \gls{ap} controls the \gls{ris} to sweep over the access codebook $\Phi_{\rm ac}$, establishing the corresponding access slots. Meanwhile, the \glspl{ue} transmit their packets according to $\Pi_k$. Let $\mc{K}_{a,n}\subseteq\mc{K}_a$ denote the subset of contending \glspl{ue} having chosen to transmit in the $n$-th access slot, \emph{i.e.}, $\mc{K}_{a,n}=\{k:n\in\Pi_k,\forall k\in\mc{K}_a\}$, $n\in\mc{N}_{\rm ac}$. The received signal $\mathbf{v}[n]\in\mathbb{C}^{L}$ at the \gls{ap} is $\mathbf{v}[n]=\sqrt{\rho_k}\sum_{k\in\mc{K}_{n}}{\zeta\UL_{a,k}[n]}\boldsymbol{\nu}_k + \boldsymbol{\eta}_a[n],$ where $\boldsymbol{\nu}_k\in\mathbb{C}^{L}$ is the packet of the $k$-th UE with zero mean and $\mathbb{E}\{\lVert\boldsymbol{\nu}_k\rVert^2_2\}=L$, and $\boldsymbol{\eta}_b[n]\in\mathbb{C}^{L}\sim\mc{N}_{\mathbb{C}}(\mathbf{0},\sigma^2\mathbf{I}_{L_{\rm ac}})$ is the receiver noise at the \gls{ap}. The \gls{ap} can run a decoding process over $\{\mathbf{v}[n]\}_{n\in\mathcal{N}_{\rm ac}}$ that can make use of collision resolution strategies, such as the one detailed in \cite[Algo. 1]{Croisfelt2022}. After the decoding is complete, the \gls{ap} holds the set of \glspl{ue} that have their frame successfully decoded as $\mc{K}_{\rm ac}\subseteq\mc{K}$ with $|\mc{K}_{\rm ac}|=K_{\rm ac}$. A set $\mc{N}_{{\rm ac},k}\subseteq\mc{N}_{\rm ac}$ is also stored containing the access slots in which each member of $\mc{K}_{\rm ac}$ was successfully decoded w/ $|\mc{N}_{{\rm ac},k}|=N_{\rm F}$. 

\section{Practical Details}\label{sec:practical}
In this section, we discuss some implementation issues of the proposed \gls{ris}-assisted \gls{ra} protocol. We start by discussing the overhead and computational complexity of the proposed protocol by focusing on the \glspl{ue}' point of view. Then, we discuss a method to implement the \gls{ris}-assisted \gls{arq} protocol, so that the \glspl{ue} can be acknowledged that their \gls{dlc} frame was successfully received by the \gls{ap}. Finally, we discuss possible ways to extend the proposed protocol for more realistic settings and considerations regarding the channel model.

\subsection{Computational Complexity and Overhead}\label{sec:practical:overhead}
Regarding computational complexity, the most expensive task on the \gls{ue} side is the learning of the $\hat{f}_k$ model. For learning, we considered interpolation methods due to their mathematical tractability, which allows us to calculate theoretical error limits. In the general case, polynomial interpolation methods have a time complexity of $O(N_{\rm co}^2)$~\cite{Hamming1986}, where $N_{\rm co}$ is the number of sampling points. As seen in Section \ref{sec:training:configuration-codebook}, $N_{\rm co}$ often ranges from tens to hundreds, resulting in computational complexity in the order of $10^2$-$10^4$. This computational complexity would definitely increase if we drop Assumptions \ref{assu:ideal} and \ref{assu:position}, since we have more degrees of freedom to be explored. In this case, \glspl{ue} could use machine learning methods to learn $\hat{f}_k$, which would require a much larger number of samples (configurations) $N_{\rm co}$. However, the relevance of this computational cost depends on the frequency in which the channel oracle is performed. Likewise, the main source of overhead in the proposed protocol is the realization of the channel oracle, whose impact also depends on how many times it is realized. As seen in the goodput in \eqref{eq:goodput}, the parameter $\alpha$ models the ratio between the number of access periods per channel oracle realization. Therefore, the impact of overhead and computational complexity are dependent on the frequency at that the channel oracle channel is performed, which in turn is related to how fast \glspl{ue} change their positions and the channel dynamics.

On one hand, in static or low mobility scenarios, the model $\hat{f}_k$ does not need to be frequently retrained and it can be reused many times, making the computational complexity and the overhead negligible, with $\alpha$ ranging from $10^{-6}$ to $10^{-2}$ ($10^{6}$ access periods/channel oracle to 100 access periods/channel oracle). This scenario works well for low-cost \glspl{ue} with limited energy, such as sensors and actuators. On the other hand, in high mobility scenarios, it may be the case that the channel oracle must be redone several times if the \glspl{ue}' position changes significantly; consequently, $\alpha$ becomes closer to 1 (1 access period/channel oracle). The cost of learning then becomes considerable, not and may not be feasible for low-cost \glspl{ue}. For more dynamic scenarios, one option would be to use similar ideas from the protocol presented here and combine them with other more dynamic learning approaches, such as reinforcement learning, if the system can be modeled as a Markov decision process.

\subsection{Frame Acknowledgments}\label{sec:practical:ack}
Here, we propose how the \gls{ris} can assist the frame-\gls{ack} process carried out by the \gls{arq} protocol. The main idea is that the \gls{ap} can design a third configuration codebook $\Phi_{\rm ack}$ to send the \gls{ack} messages based on the successfully decoded \glspl{ue} in $\mc{K}_{\rm ac}$. The goal of designing $\Phi_{\rm ack}$ is to increase the probability that the \glspl{ue} in $\mc{K}_{\rm ac}$ will be correctly informed that their messages were decoded by the \gls{ap}. To conduct the \gls{arq} protocol, the \gls{ap} simply controls the \gls{ris} to sweep over $\Phi_{\rm ack}$. For simplicity, we assume that the \gls{arq} just occurs in one round for each \gls{ue} in $\mc{K}_{\rm ac}$ and that it either fails or succeeds; it fails if the received \gls{ack} \gls{snr} at a specific \gls{ue} in $\mc{K}_{\rm ac}$ is less than a \textit{threshold \gls{snr}} $\gamma_{\rm ack}\in\mathbb{R}_{+}$. In the case of failure, the \gls{ue} drops the current frame. Below, we devise two heuristic approaches for designing $\Phi_{\rm ack}$. \textbf{Precoding-Based Acknowledgment.} Based on the maximum-ratio precoding \cite{massivemimobook}, we consider a codebook that has a single configuration: $\Phi_{\rm ack} = \{(h\DL)^{-1}({1}/{(K N_F)}\sum_{k\in\mathcal{K}_{a}}\sum_{n\in\mc{N}_{\rm ac,k}}\theta\DL_{r}[n])\},$ where we use the definition of $h\DL$ in \eqref{eq:system-model:optimal-config}. In fact, by the law of large numbers, this configuration should reflect the incoming wave toward $(\theta\DL_r\in\Phi_{\rm ack})\xrightarrow[]{}\frac{\pi}{4}$. The advantage of using a single configuration is the overhead reduction concerning the switching time $T_{\rm sw}$. \textbf{Scheduled-Based Acknowledgment.} Based on channel reciprocity, another approach is
$\Theta_{\rm ack} = \{ {1}/{N_F} \sum_{n\in\mc{N}_{\rm ac,k}} \theta\DL_r[n], \forall k\in\mc{K}_{\rm ac} \}, \text{ and } \Phi_{\rm ack} = (h\DL)^{-1}(\Theta_{\rm ack}),$
where $|\Phi_{\rm ack}|=K_{\rm ac}$ configurations. In other words, we use the average configurations associated with the access slots that led to successfully decoded packets for each \gls{ue}. We now have the opposite trade-off from before.

\subsection{Possible Extensions} \label{sec:practical:extensions}
In current wireless networks, multiple-antenna \glspl{ap} are common. In this case, it is possible to use precoding capabilities to maximize the energy transmitted/received to/from the \gls{ris} direction. Consequently, it is expected that the average \gls{snr} per \gls{ue} increases, which would eventually improve the performance of the proposed protocol. However, our main interest here is in evaluating the impact of the \gls{ris} over the protocol performance, justifying our assumption of a single-antenna \gls{ap}.

Another possibility brought by multi-antenna \glspl{ap} is the exploitation of spatial diversity to serve other devices in coverage while concurrently performing the \gls{ra} protocol. During the channel oracle, the \gls{ap} might send data toward devices in coverage taking care of neglecting the interference generated toward the \gls{ris} by means of, \textit{e.g.}, the zero-forcing precoder. During the access, \gls{ul} transmission can occur; the \gls{ap} could optimize the combining matrix in order to separate the data streams coming from the \gls{ris} and the other devices in coverage. It is worth mentioning that the \gls{ap}-\gls{ris} channel knowledge is needed to perform precoding, and, thus, the estimation of such channel needs to be performed. Fortunately, the \gls{ap} and the \gls{ris} are static and the channel between them generally remains constant over a long-time horizon~\cite{yuan2022tensor}, reducing the periodicity of \gls{chest} procedures. Nevertheless, a process orchestrating the coexistence of the proposed protocol and the communication with other \glspl{ue} needs to be designed.

Another extension is considering a scenario where multiple \glspl{ris} are deployed. We can divide the implication of applying the proposed protocol in this scenario in two: 1) the \gls{ap} controls all the \glspl{ris} to provide connectivity to the \glspl{ue} in the area of interest, and 2) the \gls{ap} controls only a subset of the \glspl{ris} in the area while a superset of them is used to serve all the \glspl{ue} in the area. In case 1), the \gls{ap} can control the configurations of the \glspl{ris} to its advantage, \emph{i.e.}, to maintain a stable behavior of the wireless environment for each configuration while avoiding interference among the signals reflected by the multiple \glspl{ris}. This case requires a specific design of the channel oracle and access codebooks taking into account that the configuration loaded by each \gls{ris} influences the equivalent channel seen by the \glspl{ue}. In case 2), the \gls{ap} cannot control the behavior of the other \glspl{ris} and, hence, the operation of the oracle might be affected and significant interference might occur. To tackle this case, a design of the channel oracle module able to minimize the impact of the interference coming from other sources might be a solution. Nevertheless, this problem might be better addressed by an orchestration between the \gls{ap} and the entities controlling the other \glspl{ris} to let the \gls{ra} protocol work when the uncontrolled \glspl{ris} do not change configurations, \emph{i.e.}, when the wireless environment is stable.

Now we discuss what would happen if we drop some of the assumptions made in Sect.~\ref{sec:system-model}.

When Assumption~\ref{assu:ideal} is dropped, the channel coefficient of the \gls{ris} cannot be expressed by simple analytical functions. The same protocol can be applied, taking care of handling the increased complexity of the design of the codebook. The access codebook can be obtained by the use of pre-defined \gls{ris} configurations pointing toward different directions, usually stored in a lookup table. Therefore, the design should only focus on finding a sufficient number of configurations to cover the area of interest. However, designing the channel oracle codebook is trickier due to the need of learning the model of the channel coefficient for all the possible reflection angles. This is an interesting learning problem that can be tackled by finding the minimum subset of the access codebook that allows an accurate estimation of $\hat{f}_k$. On the other hand, given a channel oracle codebook, different machine learning techniques can be used to approximate $\hat{f}_k$. This problem can be addressed in future works.

If Assumption~\ref{assu:position} is removed, the \glspl{ue}, the \gls{ap}, and the \gls{ris} are placed on different planes. In this case, both the channel oracle and access codebooks would need to change to account for the increased dimensions of the problem. To design the channel oracle codebook, we can use the generalization of the Nyquist-Shannon theorem in multi-dimensional spaces to obtain the \emph{lattice} of points in the azimuth and elevation angles space that assures the reconstruction of the channel coefficient~\cite{PETERSEN1962279,Kunsch2005}. This lattice of points represents the reflection angles of the configurations of the codebook that allow us to learn the model $\hat{f}_k$. Similarly, the access codebook design would need to account for different elevation angles. Nevertheless, the procedure described in Sect.~\ref{sec:access:codebook} can be easily extended to the two-dimension space, considering that a 3D half-power beamwidth can be approximated by an elliptic cone~\cite{Balanis2012antenna}.

Finally, we remark that the proposed \gls{ra} protocol can help develop new and more practical \gls{chest} and localization methods for \gls{ris}-assisted systems. It is crucial to observe that the access policies often encourage the choice of slots that are related to configurations that are in its turn correlated with the position of the \glspl{ue}. Such prior knowledge could be useful to improve such methods in practical systems, having in view the vast literature on \gls{chest} motivated by the objective of decreasing its computational complexity~\cite{Wei2021ce,yuan2022tensor,yuan2021frequency,wang2020compressed}. 
\section{Numerical Results}\label{sec:results}

\begin{table}[t]
    \caption{Simulation Parameters}
    \label{tab:simulation-parameters}
    \centering
    \resizebox{1\columnwidth}{!}{%
    \begin{tabular}{lr|lr}
        \textbf{Parameter} & \textbf{Value} & \textbf{Parameter} & \textbf{Value} \\  
        \hline
        \rowcolor[HTML]{EFEFEF}
        carrier frequency, $f_c$ & 3 GHz & antenna gains, $G_a,G_k$ & $5$ dBi \\
        num. of elements along axes, $M_x$, $M_z$ & 10 & AP transmit power, $\rho_a$ & $20$ dBm \\
        \rowcolor[HTML]{EFEFEF}
        element sizes, $d_x$, $d_z$ & $\lambda$ & UE transmit power, $\rho_k$ & $10$ dBm \\
        max. and min distances, $d_{\max}$, $d_{\min}$ & 20, 5 m & noise power, $\sigma^2$ & -94 dBm \\
        \rowcolor[HTML]{EFEFEF}
        AP-RIS distance, $d_a$ & $d_{\min}$ & threshold decoding SNRs, $\gamma_{\rm ac},\gamma_{\rm ack}$ & $3$ dB\\
        AP-RIS angle, $\theta_a$ & 45$^\circ$ & num. of frames, symbs., and reps. $N_{F},L, R$ & $1$
    \end{tabular}
    }
\end{table}

In this section, we evaluate the effectiveness of the proposed protocol.\footnote{The code to reproduce the figures is available online on \url{https://github.com/victorcroisfelt/ris-random-access-channel-oracle}.} Table \ref{tab:simulation-parameters} summarizes the standard simulation parameters used. To reduce the impact of the pathloss from the \gls{ap}-\gls{ris} link and increase as much as possible the channel gain experienced by the \glspl{ue}, we place the \gls{ap} at the minimum distance $d_{\min}$ that satisfies the far-field requirements and place it onto the bisector of the first quadrant of the system setup depicted in Fig.~\ref{fig:system-setup}, such that $\theta_a=45^{\circ}$. Our goal in positioning the \gls{ap} in this way is to emphasize how the difference in the distances from the \glspl{ue} to the \gls{ris} influences the protocol performance. As a point of reference in comparing different access policies, we take the simplest case possible where each \gls{ue} sends a single copy of each packet $R=1$, except for the SMAP with $R\leq2$. For the same reason, we consider that a frame comprises a single packet and a packet comprises a single symbol, meaning that $N_{F}=L=1$. Considering the industrial shed example (Sect.~\ref{sec:system-model}), we assume that \glspl{ue} are distributed according to the \glspl{pdf} in \eqref{eq:pdfs} with a maximum distance of $20$ meters. With the transmit and noise power from Table \ref{tab:simulation-parameters}, the \gls{dl} received SNR ranges from approximately -108 to 36 dB, while -118 to 26 dB is the range for the \gls{ul} received SNR. The median and average values for the \gls{ul} received SNR are approximately $1.50$ and $2.10$ dB. Based on such values, we choose the threshold decoding \gls{snr} as 3 dB.\footnote{The main objective of the numerical results shown here is to evaluate the gains obtained with the proposed protocol in the worst possible conditions in terms of the \gls{dlc} frame design. If the protocol surpasses the baseline performance under these conditions, we expect that by further optimizing other parameters, the protocol will have even greater gains.} Finally, to model the unpredictability of active \glspl{ue}, we consider that $K_a$ is Poisson distributed with parameter $\kappa$ being the \emph{channel load}.

\noindent \textbf{Setting Parameters.} For the proposed protocol, we need to set up the following parameters: number of channel oracle configurations (samples) $N_{\rm co}$, pilots length $L_{\rm co}$, and access period $N_{\rm ac}$. The selection of the first two parameters is studied and done in Section \ref{sec:results:channel-oracle}. For $N_{\rm ac}$, we assume that the \gls{ap} knows the channel load by using some estimation over time and set $N_{\rm ac}=\kappa$. Moreover, spline interpolation is used by the \glspl{ue} to obtain $\hat{f}_k$. For simplicity, we evaluate a very static scenario such that the throughput in \eqref{eq:goodput} is equal to the goodput in \eqref{eq:goodput} being $\alpha\approx0$.

\noindent \textbf{Baseline.} As a baseline, we consider the legacy \gls{saloha}, which does not benefit from the \gls{ris}. Each \gls{ue} selects a slot uniformly at random without replacement from $N_{\rm ac}=\kappa$. Consequently, the channel oracle module is ignored and the throughput equals the goodput (see eq.~\eqref{eq:goodput}).

\subsection{Channel Oracle} \label{sec:results:channel-oracle}
In this part, we study and select $N_{\rm co}$ and $L_{\rm co}$. We start by evaluating how good is the procedure developed in Section \ref{sec:channel-oracle} to obtain $\hat{f}_k$. Fig.~\ref{fig:results:training:performance:se} shows the normalized expected \gls{se} of the model as specified in Corollary~\ref{cor:training:reconstruction} when considering different estimation tolerances $\delta\DL_{\rm tol}$ defined in Corollary~\ref{cor:training:estimation}. The figure evaluates the design of the configuration codebook carried out in Sect.~\ref{sec:training:configuration-codebook} by vertically drawing some of the approximated lower bounds obtained in Remark \ref{remark:n-co-values}. As expected from the result of Corollary~\ref{cor:training:reconstruction}, we verify that the reconstruction error is dominated by noise and estimation effects parameterized by $\delta\DL_{\rm tol}$. For the most conservative bound of ${N}_{\rm co}=16$ (median w/ $\epsilon=10^{-2}$), the expected \gls{se} is considerably high on the order of $10^{-1}$, showing that this bound fairly undersamples the function we are interested in reconstructing. On the other hand, both ${N}_{\rm co}=142$ (maximum w/ $\epsilon=10^{-2}$) and ${N}_{\rm co}=150$ (Taylor-approximation) oversample the function since they ensure the same quality that ${N}_{\rm co}=46$ (median w/ $\epsilon=10^{-3}$) does. \textit{Thus, we set ${N}_{\rm co}$ according to the approximated bound of ${N}_{\rm co}=46$ configurations because it provides a good compromise between the overhead $T_{\rm co}$ and the reconstruction error.} For the choice of $L_{\rm co}$, we have observed through simulations that an error tolerance $\delta\DL_{\rm tol}=10^{-3}$ implies a good $\overline{\mathrm{SE}}$, according to Corollary~\ref{cor:training:reconstruction}. Thus, we set $L_{\rm co}=1$ according to~\eqref{eq:training:dl-received:channel-uses-bound}.

\begin{figure}[t]
    \centering
    \vspace{-.5mm}
    \input{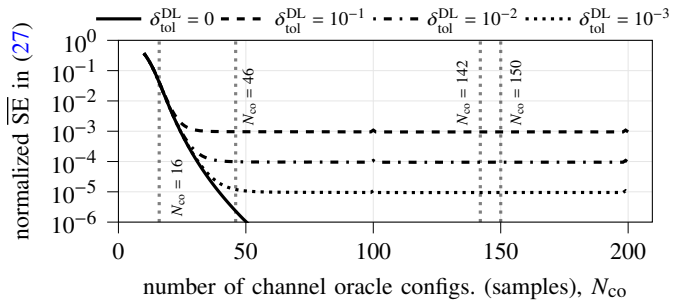}
    \vspace{-6mm}
    \caption{Normalized expected squared error (SE) of the reconstructed model $\hat{\zeta}\DL_k=\hat{f}_k(\theta\DL_r)$ when using spline interpolation function with different estimation tolerances $\delta\DL_{\rm tol}$. Noiseless reconstruction happens when $\delta\DL_{\rm tol}=0$, which shows the error inflicted by the interpolation process. Vertical dotted \textcolor{gray}{gray} lines represent different choices of $N_{\rm co}\in\{16, 46, 142, 150\}$ given in Remark~\ref{remark:n-co-values}.}
    \label{fig:results:training:performance:se}
\end{figure}

\subsection{Throughput and Impact of RIS Hardware}
Fig.~\ref{fig:results:throughput} evaluates the expected overall throughput $\overline{\mathrm{TP}}$ given in~\eqref{eq:goodput} for the proposed \gls{ris}-assisted protocol considering the three different access policies with no switching time, $T_{\rm sw}=0$. Disregarding access policy, our proposed random access protocol always outperforms the baseline. On average, our best results are obtained by the $R$-GSCAP access policy that provides a throughput 66.18\% higher than the baseline. In Fig.~\ref{fig:results:switching-time}, we evaluate how the throughput is impacted by the control commands and the hardware operation at the \gls{ris}. One can note that if the switching time is the same duration as the symbol time, the protocol loses its practicality due to a large overhead from sweeping over the access configurations. Thus, from a protocol point of view, we would like to have fast-switching \glspl{ris} and a fast \gls{cc} between \gls{ap} and \gls{ris}. This \gls{cc} and hardware dependencies comprise one of the major disadvantages of the proposed protocol. One way to reduce the impact of the switching time could be to reduce the size of the access codebook at the cost of more collisions on average. 

\begin{figure}[t]
    \centering
    \vspace{-3mm}
    \input{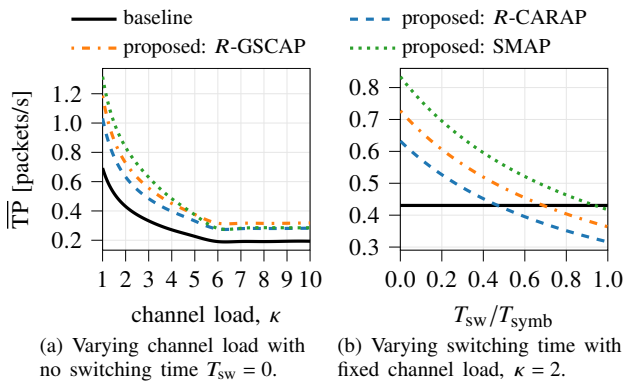}
    \vspace{-6mm}
    \caption{Expected overall throughput, $\overline{\mathrm{TP}}$, in \eqref{eq:goodput} when assuming that $N_{\rm ac}=\kappa$ and $N_{\rm ac}$ always respects the bound in~\eqref{eq:access:lower-bound-access-slots}.}
    \label{fig:results:performance}
\end{figure}
\begin{figure}[ht]
    \centering
    \input{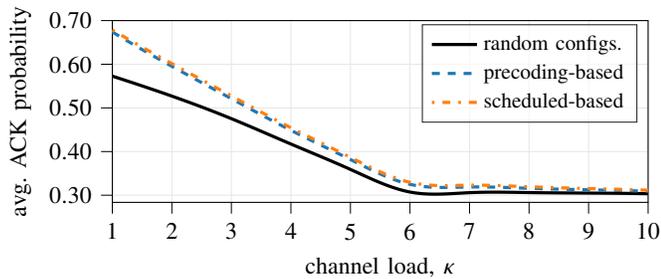}
    \vspace{-6mm}
    \caption{Evaluation of \gls{ris}-assisted frame \gls{ack} strategies when considering the proposed protocol using the $R$-GSCAP (best average performance in Fig.~\ref{fig:results:performance}).}
    \label{fig:results:dlc-ack}
\end{figure}

\subsection{Frame Acknowledgments}
Fig.~\ref{fig:results:dlc-ack} shows the average probability of \gls{ack} when using random configurations, precoding-based, and scheduled-based \gls{ris}-assisted frame \gls{ack} strategies. The latter two were proposed in Section~\ref{sec:practical:ack}, where the first is a baseline scheme in which a random configuration is loaded at the \gls{ris} when the \gls{ap} is going to \gls{ack} the packet of a \gls{ue}. One can see that the proposed strategies perform better than the baseline \gls{ack} scheme, showing that it is advantageous to exploit information obtained during the access to communicate back with the \glspl{ue}.

\section{Conclusion}\label{sec:conclusions}
We proposed a new \gls{ris}-assisted \gls{ra} protocol. It carefully introduces the \gls{ris} into the \gls{mac} layer, exploiting its \gls{phy} capabilities to create coordination in the transmission of uncoordinated \glspl{ue} when considering the most challenging case that no information about the \glspl{ue} is available at the \gls{ap}. In our experiments, we showed that our protocol can outperform the legacy \gls{saloha} by approximately 60\% on average. However, our protocol is highly dependent on the quality of the \gls{cc} between the \gls{ap} and the \gls{ris}, the \gls{ris} hardware, and the mobility of the \glspl{ue}. This work opens up several new research directions to extend the current protocol and propose new ideas on how to integrate RIS into higher-layer protocols.



\bibliographystyle{IEEEtran}

\end{document}